\documentclass{aa}
\usepackage{graphicx,latexsym}
\begin{document}

\title{Geometrical Constraints on Dark Energy}

\author{A. K. D. Evans~\inst{1}
\and I. K. Wehus~\inst{2}
\and {\O}. Gr\o n~\inst{3,2}
\and {\O}. Elgar{\o}y~\inst{1}
}

\institute{
Institute of Theoretical Astrophysics, University of Oslo, P.O. Box 1029, Blindern, N-0315 Oslo, Norway 
\and Department of Physics, University of Oslo, P. O. Box 1048, Blindern, N-0316 Oslo, Norway 
\and Oslo College, Faculty of Engineering, Cort Adelers gt. 30, N-0254 Oslo, Norway 
}

\offprints{oelgaroy@astro.uio.no}

\date{Received/Accepted}

\abstract{{
We explore the recently introduced statefinder parameters.     
After reviewing their basic 
properties, we calculate the statefinder parameters for a variety of 
cosmological models, and investigate their usefulness as a means of 
theoretical classification of dark energy models.  
We then go on to consider their use in 
obtaining constraints on  dark energy from present and future    
supernovae type Ia data sets.  Provided that the statefinder parameters 
can be extracted unambiguously from the data, they give a good visual 
impression of where the correct model should lie.  However, it is 
non-trivial to extract the statefinders from the data in a model-independent 
way, and one of our results indicates that an expansion of the 
dark energy density to second order in the redshift is inadequate.  
Hence, while a useful theoretical and visual tool, 
applying the statefinders to observations is not straightforward.  
}
\keywords{Cosmology:theory -- cosmological parameters}
}

\authorrunning{Evans et al.}

\maketitle

\section{Introduction}

It is generally accepted that we live in an accelerating universe. 
Early indications of this fact  
came from the magnitude-redshift relationship of galaxies (Solheim 1966), 
but the reality 
of cosmic acceleration was not taken seriously until the 
magnitude-redshift relationship was measured recently using high-redshift 
supernovae type Ia (SNIa) as standard candles 
(Riess et al. 1998, Perlmutter et al. 1999).  
The observations can be explained by invoking 
a contribution to the energy density with negative pressure, 
the simplest possibility being Lorentz Invariant Vacuum Energy (LIVE), 
represented by a cosmological constant.  Independent 
evidence for a non-standard contribution to the energy budget of 
the universe comes from e.g. the combination of the power spectrum of the 
cosmic microwave background (CMB) temperature anisotropies and large-scale 
structure:  the position of the first peak in the CMB power spectrum 
is consistent with the universe having zero spatial curvature, 
which means that 
the energy density is equal to the critical density.  However, 
several probes of the large-scale matter distribution show 
that the contribution of standard 
sources of energy density, whether luminous or dark, is only a 
fraction of the critical density.  Thus, an extra, unknown component 
is needed to explain the observations (Efstathiou et al. 2002; 
Tegmark et al. 2004).

Several models describing an accelerated universe have been suggested.  
Typically, they are tested against the SNIa data on a model-by-model 
basis using the relationship between luminosity distance and redshift, 
$d_{\rm L}(z)$,  
defined by the model.  Another popular approach is to parametrize classes of 
dark energy models by their prediction for the so-called equation of 
state $w(z) \equiv p_{\rm x}/\rho_{\rm x}$, where $p_{\rm x}$ and 
$\rho_{\rm x}$ are the pressure and the energy density, respectively, of 
the dark energy component in the model.  One can then Taylor expand 
$w(z)$ around $z=0$.  The current data allow only relatively weak 
constraints on the zeroth-order term $w_0$ to be derived.  
A problem with this approach is that some attempts at explaining the 
accelerating Universe do not involve a dark component at all, but 
rather propose modifications of the Friedmann equations   
(Deffayet 2001; Deffayet, Dvali \& Gabadadze 2002; Dvali, Gabadadze \& Porrati 2000; Freese \& Lewis 2002; Gondolo \& Freese 2003; Sahni \& Shtanov 2003).  
Furthermore, it is possible for two different dark energy models to 
give the same equation of state, as discussed by \cite{paddy1} 
and \cite{paddy2}.

Recently, an alternative way of classifying dark energy models using 
geometrical quantities was proposed (Sahni et al. 2003, Alam et al. 2003).  
These so-called 
statefinder parameters are constructed from the Hubble parameter $H(z)$ 
and its derivatives, and in order to extract these quantities in a 
model-independent way from the data, one has to parametrize $H$ in 
an appropriate way.  This approach was investigated at length in 
\cite{alam} using simulated data from a SNAP\footnote{see http://snap.lbl.gov}-type experiment.  
In this paper, we present a further investigation of this formalism.  
We generalize the formalism to universe models with 
spatial curvature in Section 2, and give expressions for the statefinder 
parameters in several specific dark energy models.   
In the same section, we also take 
a detailed look at how the statefinder parameters behave for quintessence 
models, and show that some of the statements about these models 
in \cite{alam} have to be modified.  In Section 3 we discuss what can 
be learned from current SNIa data, considering both direct $\chi^2$ fitting 
of model parameters to data, and statefinder parameters.  In Section 4 
we look at simulated data from an idealized SNIa survey, showing that 
reconstruction of the statefinder parameters from data is likely to 
be non-trivial. Finally, Section 5 contains our conclusions.

\section{Statefinder parameters: definitions and properties}

The Friedmann-Robertson-Walker models of the universe have earlier been 
characterized by the Hubble parameter and the deceleration parameter, 
which depend on the first and second derivatives of the scale factor, 
respectively: 
\begin{eqnarray}
H &= & \frac{\dot{a}}{a} \label{eq:1.1} \\
q & =& -\frac{\ddot{a}a}{\dot{a}^2}=-\frac{\dot H}{H^2}-1, \label{eq:eq1.2}
\end{eqnarray}
where dots denote differentiation with respect to time $t$.   
The proposed SNAP satellite will provide accurate determinations of the 
luminosity distance and redshift of more than 2000 supernovae of type Ia.  
These data will permit a very 
precise determination of $a(z)$.  It will then be important to include 
also the third derivative of the scale factor in our characterization of 
different universe models.

Sahni and coworkers (Sahni et al. 2003, Alam et al. 2003) recently 
proposed a new pair of 
parameters $(r,s)$ called {\it statefinders} as a means of  
distinguishing between different types of dark energy.  
The statefinders were introduced to characterize flat universe 
models with cold matter (dust) and dark energy.  They were defined as
\begin{eqnarray}
r & = & \frac{\stackrel{\bf{...}}{a}}{aH^3} 
=\frac{\ddot H}{H^3}+3\frac{\dot H}{H^2}+1\label{eq:eq1.3} \\
s & = & \frac{r-1}{3\left(q-\frac{1}{2}\right)}. \label{eq:eq1.4}
\end{eqnarray}
Introducing the cosmic redshift $1+z = 1/a \equiv x$,  
we have $\dot{H} = -H'H/a$, where $H' = dH/dx$, the deceleration parameter 
is given by 
\begin{equation}
q(x) = \frac{H'}{H}x-1.
\label{eq:eq1.9}
\end{equation}
Calculating $r$, making use of $a'=-a^2$, we obtain 
\begin{equation}
r(x) = 1-2\frac{H'}{H}x + \left(\frac{H'^2}{H^2}+\frac{H''}{H}\right)x^2.
\label{eq:eq1.14}
\end{equation}
The statefinder $s(x)$, for flat universe models, is then found by inserting the expressions 
(\ref{eq:eq1.9}) and (\ref{eq:eq1.14}) into equation (\ref{eq:eq1.4}).  
The generalization to non-flat models will be given in the next subsection.  

The Friedmann equation takes the form\footnote{Throughout this paper we use units where the speed of light $c=1$.} 
\begin{equation}
H^2 = \frac{8\pi G}{3}(\rho_{\rm m} + \rho_{\rm x}) -\frac{k}{a^2},
\label{eq:eq1.5}
\end{equation}
where $\rho_{\rm m}$ is the density of cold matter and $\rho_{\rm x}$ is 
the density of the dark energy, and $k=-1,0,1$ is the curvature parameter 
with $k=0$ corresponding to a spatially flat universe.   
The dust component is pressureless, so 
the equation of energy conservation implies 
\begin{equation}
\rho_{\rm m}  =  \rho_{\rm m 0} a^{-3}.  
\label{eq:eq1.6} 
\end{equation}
This gives for the density of dark energy:
\begin{eqnarray}
\rho_{\rm x} &=& \rho_{\rm c} - \rho_{\rm m} -  \frac{3k}{8\pi G a^2}
\nonumber \\ 
&=& \frac{3}{8\pi G}(H^2 +k x^2-\Omega_{\rm m 0}H_0^2 x^3),
\label{eq:eq1.10}
\end{eqnarray}
where 
and $\Omega_{\rm m 0}$ and $\Omega_{\rm x 0}$ are the 
present densities of matter and dark energy, respectively, in units 
of the present critical density $\rho_{\rm c 0}= 3H_0^2 / 8\pi G$.  
In the following, we will use the notation $\Omega_{\rm i}
\equiv 8\pi G \rho_{\rm i}(t)/H^2(t)$, $\Omega_{\rm i 0} \equiv 
\Omega_{\rm i}(t=t_0)$, where $t_0$ is the present age of the Universe,
and also $\Omega=\sum_{\rm i}\Omega_{\rm i}$. 
From Friedmann's acceleration equation 
\begin{equation}
\frac{\ddot{a}}{a} = -\frac{4\pi G}{3}\sum_{\rm i} (\rho_{\rm i}+
3p_{\rm i}),
\label{eq:eq1.11}
\end{equation}
where $p_{\rm i}$ is the contribution to the pressure from component i, 
it follows that 
\begin{equation}
p_{\rm x} = \frac{H^2}{4\pi G}\left(q-\frac{\Omega}{2} \right) = 
\frac{3}{8\pi G}\left[\frac{1}{3}(H^2)'x -\frac{k}{3}x^2 -H^2\right].
\label{eq:eq1.12}
\end{equation}
Hence, if dark energy is described by an equation of state 
$p_{\rm x} = w(x)\rho_{\rm x}$, we have 
\begin{equation}
w(x) = \frac{\frac{1}{3}(H^2)'x-H^2-\frac{k}{3}x^2}{H^2 + k x^2 -
  H_0^2\Omega_{\rm m 0} x^3}. 
\label{eq:eq1.13}
\end{equation}

In the following subsections, we calculate statefinder parameters for 
universe models with different types of dark energy.  

\subsection{Models with an equation of state $p=w(z)\rho$}

First we consider dark energy obeying an equation of state of the 
form $p_{\rm x} = w\rho_{\rm x}$, where $w$ may be time-dependent.  
Quintessence models (Wetterich 1988; Peebles \& Ratra 1988), 
where the dark energy is provided by a scalar field evolving in time, 
fall in this category. 
The formalism in \cite{sahni} and \cite{alam} will be generalized to permit 
universe models with spatial curvature.  Then equation (\ref{eq:eq1.4}) is 
generalized to 
\begin{equation}
s = \frac{r-\Omega}{3(q-\Omega/2)},
\label{eq:eq1.145}
\end{equation}
where $\Omega = \Omega_{\rm m}+ \Omega_{\rm x}=1-\Omega_{\rm k}$, and 
$\Omega_{\rm k}=-k/(a^2H^2)$.    

The deceleration parameter can be expressed as 
\begin{equation}
q = \frac{1}{2}[\Omega_{\rm m} + (1+3w)\Omega_{\rm x}] = \frac{1}{2}(\Omega+3w\Omega_{\rm x}).
\label{eq:eq1.15}
\end{equation}
Differentiation of equation (\ref{eq:eq1.2}) together with equation 
(\ref{eq:eq1.3}) leads to 
\begin{equation}
r = 2q^2 + q -\frac{\dot{q}}{H}.
\label{eq:eq1.16}
\end{equation}
From equation (\ref{eq:eq1.15}) we have 
\begin{equation}
\dot{q} = \frac{1}{2}\dot{\Omega}_{\rm m}+\frac{1}{2}(1+3w)
\dot{\Omega}_{\rm x} + \frac{3}{2}\dot{w}\Omega_{\rm X}. 
\label{eq:eq1.17}
\end{equation}
Furthermore, 
\begin{equation}
\dot{\Omega} = \frac{\dot{\rho}}{\rho_{\rm c}}-\frac{\rho}{\rho_{\rm c}^2}
\dot{\rho}_{\rm c},
\label{eq:eq1.18}
\end{equation}
with 
\begin{equation}
\dot{\rho}_{\rm c} = \frac{3H\dot{H}}{4\pi G}, 
\label{eq:eq1.19}
\end{equation}
and 
\begin{equation}
\dot{H} = -H^2(1+q),
\label{eq:eq1.20}
\end{equation}
giving 
\begin{equation}
\dot{\rho}_{\rm c} = -2(1+q)H\rho_{\rm c},
\label{eq:eq1.21}
\end{equation}
which leads to 
\begin{equation}
\dot{\Omega} = \frac{\dot{\rho}}{\rho_{\rm c}} + 
2(1+q)H\Omega.
\label{eq:eq1.22}
\end{equation}
For cold matter, $\dot{\rho}_{\rm m} = -3H\rho_{\rm m}$, giving 
\begin{equation}
\dot{\Omega}_{\rm m} = (2q-1)H\Omega_{\rm m}, 
\label{eq:eq1.23}
\end{equation}
and for the dark energy, $\dot{\rho}_{\rm x}=-3(1+w)H\rho_{\rm x}$, giving 
\begin{equation}
\dot{\Omega}_{\rm x} = (2q-1-3w)H\Omega_{\rm x}.
\label{eq:eq1.24}
\end{equation}
Inserting equations (\ref{eq:eq1.23}) and (\ref{eq:eq1.24}) into equation 
(\ref{eq:eq1.17}) and the resulting expression into (\ref{eq:eq1.16}) 
finally leads to 
\begin{equation}
r = \Omega_{\rm m}+\left[1+\frac{9}{2}w(1+w)\right]\Omega_{\rm x} 
-\frac{3}{2}\frac{\dot{w}}{H}\Omega_{\rm x}.
\label{eq:eq1.25}
\end{equation}
Inserting equation (\ref{eq:eq1.25}) into equation (\ref{eq:eq1.145}) gives 
\begin{equation}
s = 1 + w -\frac{1}{3}\frac{\dot{w}}{wH}.
\label{eq:eq1.27}
\end{equation}

For a flat universe $\Omega_{\rm m}+\Omega_{\rm x}=1$ and the 
expression for $r$ simplifies to 
\begin{equation}
r = 1 + \frac{9}{2}w(1+w)\Omega_{\rm X}-\frac{3}{2}\frac{\dot{w}}{H}
\Omega_{\rm x}.
\label{eq:eq1.26}
\end{equation}
Note that for the case of LIVE, $w=-1=\,{\rm constant}$, 
and one finds $r=\Omega$, $s=0$ for all redshifts.  
For a model with curvature and matter only one gets 
$r=2q=\Omega_{\rm m}$, $s=2/3$.  The same result is obtained 
for a flat model with matter and dark energy with a constant 
equation of state $w=-1/3$, which is the equation of state of  
a frustrated network of non-Abelian cosmic strings 
(Eichler 1996; Bucher \& Spergel 1999).  Thus, the statefinder parameters 
cannot distinguish between these two models.   However, neither of these 
two model universes are favoured by the current data (for one thing, 
they are both decelerating), so this is 
probably an example of academic interest only.  

\begin{figure}[htbp!]  
  \begin{center}
 \includegraphics[width=60mm,height=60mm]{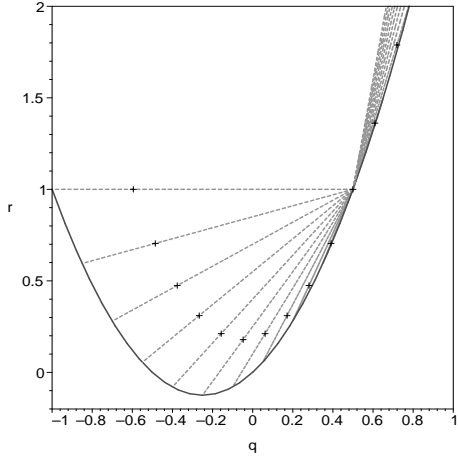}
  \end{center}
\caption{The $q-r$-plane for flat matter+quiessence models. 
The horizontal curve has $w=-1$ ($\Lambda$CDM). Then $w$
  increases by $1/10$ 
  counterclockwise until we reach $w=1$ in the upper right. 
When $\Omega_{\rm x 0}=0$ all models start at the point $q=0.5$, $r=1$
(Einstein-de Sitter model). As  $\Omega_{\rm x 0}$ increases every
model moves towards the solid curve which marks $\Omega_{\rm x 0}=1$.
The crosses mark the present epoch.} 
 \label{fig:larsfig}
\end{figure}

For a constant $w$, and $\Omega_{\rm m 0}+\Omega_{\rm x 0} = 1$, the 
$q$--$r$ plane for different values of $\Omega_{\rm x}$ and $w$
is shown in figure \ref{fig:larsfig}.  
Quintessence with $w={\rm constant}$ is called quiessence.  
The relation between $q$ and $r$ for flat universe models with
matter+quiessence  is found by eliminating $\Omega_{\rm x}$ between equation (14), 
with $\Omega=1$, and equation (26).  This gives 
\begin{equation}
r = 3(1+w)q-\frac{1}{2}(1+3w), 
\end{equation}
which is the equation of the dotted straight lines in figure \ref{fig:larsfig}.
When $\Omega_{\rm x} = 1$, all 
models lie on the solid curve given by 
\begin{eqnarray}
q&=&\frac{3}{2}w+\frac{1}{2} \\
r&=&\frac{9}{2}w(1+w)+1, 
\end{eqnarray}
or 
\begin{equation}
r = 2q^2 +q,
\end{equation}
in accordance with equation (15) since $\dot{q}=0$ for these models. 
This curve is the lower bound for all models with a constant $w$.  
For $-1 \leq w \leq 0$, all matter+quiessence models will at any time 
fall in the sector between this curve and the $r=1$-line which corresponds 
to $\Lambda{\rm CDM}$.  The results shown in Alam et al. (2003) 
seem to indicate that  all 
matter+quintessence models will fall within this same sector as the 
matter+quiessence models do.  However, as we will show below, this is 
not strictly correct.  

\subsection{Scalar field models}

If the source of the dark energy is a scalar field $\phi$, as in the 
quintessence models (Wetterich 1988; Peebles \& Ratra 1988), the equation 
of state factor $w$ is 
\begin{equation}
w = \frac{\dot{\phi}^2 - 2V(\phi)}{\dot{\phi}^2 + 2V(\phi)}.
\label{eq:eq1.28}
\end{equation}
Differentiation gives 
\begin{equation}
\dot{w}\rho_{\rm x} = \frac{2\dot{\phi}(2\ddot{\phi}V-\dot{\phi}\dot{V})}
{\dot{\phi}^2 + 2V}.
\label{eq:eq1.29}
\end{equation}
Using the equation of motion of the scalar field
\begin{equation}
\ddot{\phi} = -3H\dot{\phi}-\frac{dV}{d\phi},
\label{eq:eq1.30}
\end{equation}
and $\dot{V}=\dot{\phi}dV/d\phi$ in equation (\ref{eq:eq1.28}) and 
inserting the result in equation (\ref{eq:eq1.25}) we obtain 
\begin{equation}
r = \Omega + 12\pi G \frac{\dot{\phi}^2}{H^2}+8\pi G \frac{\dot{V}}{H^3},
\label{eq:eq1.31}
\end{equation}
and furthermore,
\begin{equation}
q -\frac{\Omega}{2}=\frac{3}{2}w\Omega_{\rm x} = 4\pi G \frac{p_{\rm x}}
{H^2} = \frac{4\pi G}{H^2}\left(\frac{1}{2}\dot{\phi}^2 - V\right).
\label{eq:eq1.32}
\end{equation}
Hence the statefinder $s$ is 
\begin{equation}
s = \frac{2\left(\dot{\phi}^2 + \frac{2}{3}\frac{\dot{V}}{H}\right)}
{\dot{\phi}^2 - 2V}.
\label{eq:eq1.33}
\end{equation}

\begin{figure}[]  
  \begin{center}
\includegraphics[width=60mm,height=60mm]{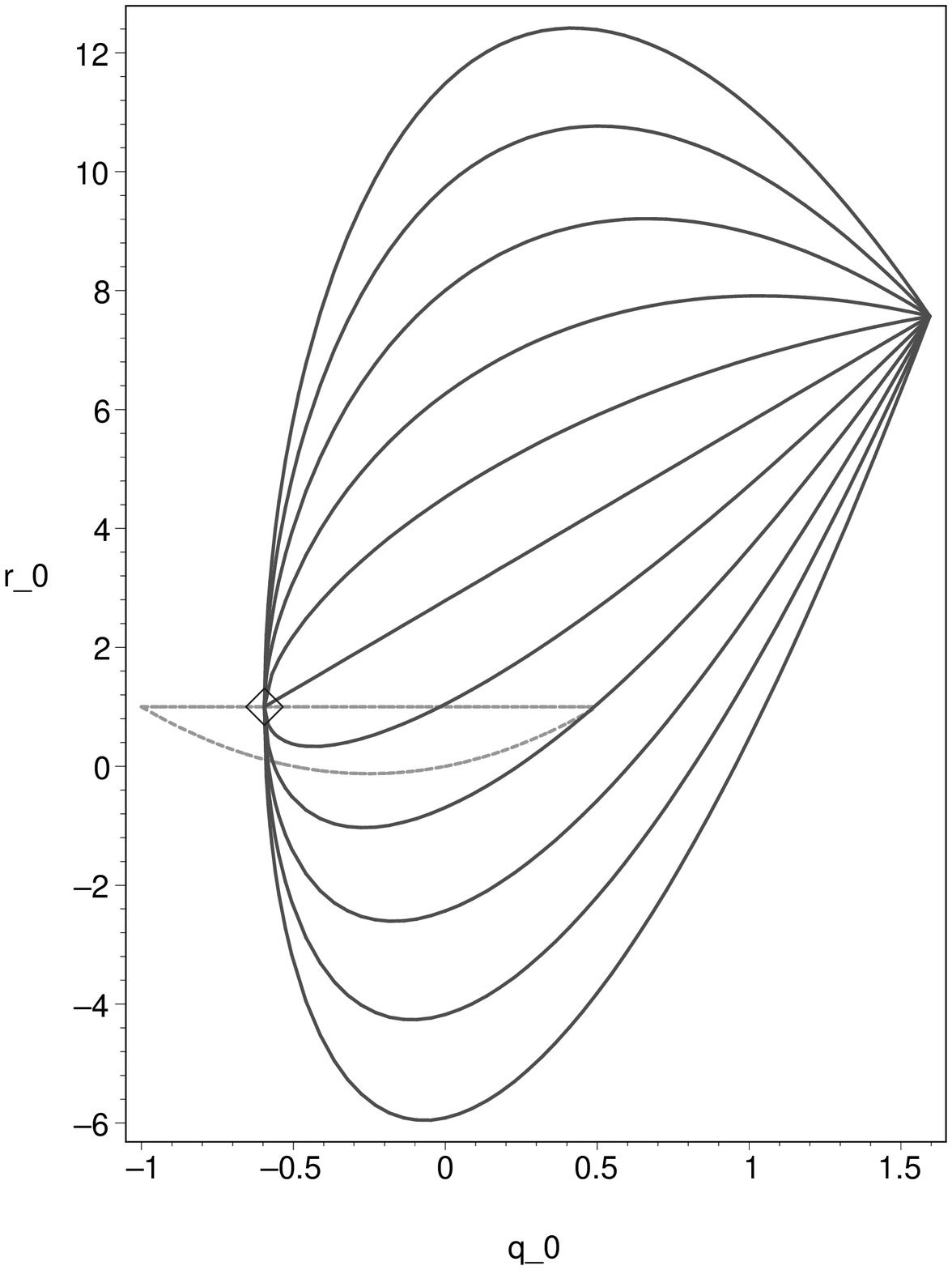}
\includegraphics[width=60mm,height=60mm]{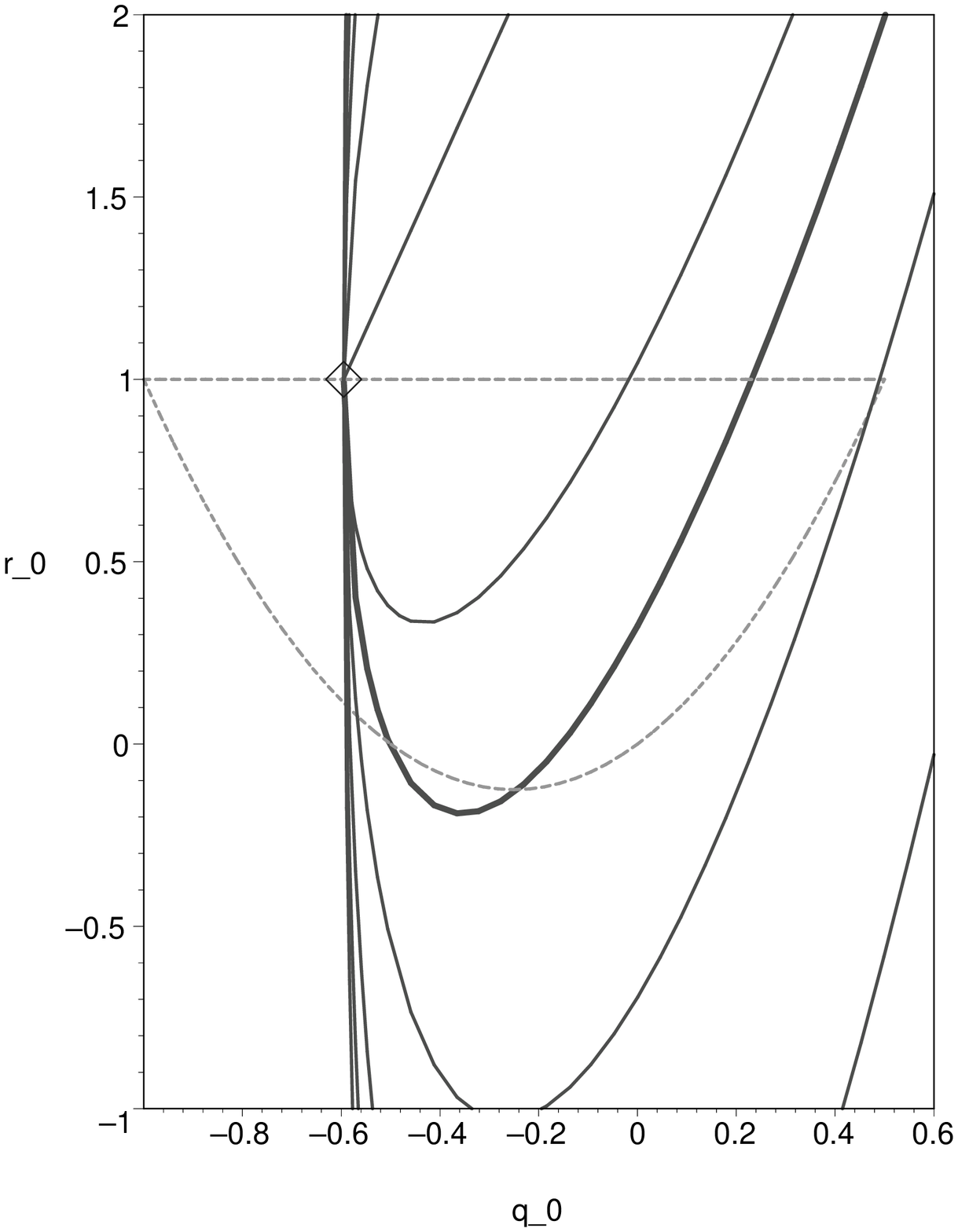}
  \end{center}
\caption{Present values of $q$ and $r$ for matter+quintessence with
  an exponential potential. \newline 
  Top panel: From top to bottom the different curves have
  $\lambda=-5,-4,-3,-2,-1,0,1,2,3,4,5$. They all start at the
  point $q_0(\Omega_{\rm kin}=.73)=1.595$,
  $r_0(\Omega_{\rm kin}=.73)=7.57)$ [matter+Zeldovich gas ($p_{\rm x}
=\rho_{\rm x}$)].  As
  $\Omega_{\rm kin}$ decreases when we move to the left, they 
  join at the point $q_0(\Omega_{\rm kin}=0)=-0.595$, $r_0(\Omega_{\rm
  kin}=0)=1)$  ($\Lambda \rm CDM$, marked with a diamond). The
  dotted curve shows  the area all matter+quiessence models must lie
  within at all times.
\newline Bottom panel: Zoom-in of the figure above. Here the curve
having $\lambda^2=2$ is also plotted (thick line). } 
 \label{r0q0}
\end{figure}

\begin{figure}[htbp!]  
  \begin{center}
\includegraphics[width=60mm,height=60mm]{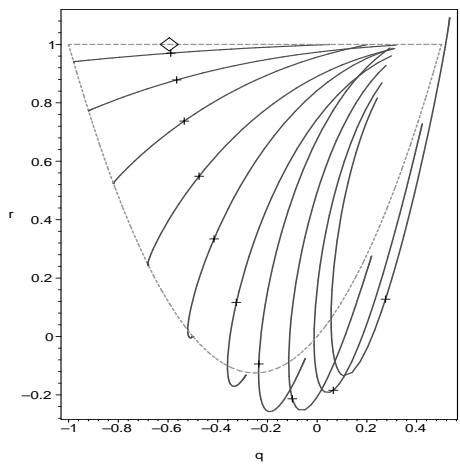}
  \end{center}
\caption{Time-evolution of $q$ and $r$ for models with matter and quintessence
  with an exponential potential. The crosses mark the present epoch.
  The diamond represents the present $\Lambda \rm CDM$ model. 
The curve on top has  $\lambda = 0.2$ and then $\lambda$ increases by
  $0.2$ for each curve going counter-clockwise until we reach
  $\lambda=2$ to the right. The corresponding values for $\Omega_{\rm
  kin}$ today are $\Omega_{\rm kin0}=0.002, 0.01, 0.02, 0.04, 0.06, 0.09,
  0.12, 0.165, 0.22, 0.29$. 
 The dotted curve
  shows the area all matter+quiessence models must lie within at all
  times. We see that all models will eventually move towards this curve.
}
 \label{rqexp}
\end{figure}

\begin{figure}[htbp!]  
  \begin{center}
\includegraphics[width=60mm,height=60mm]{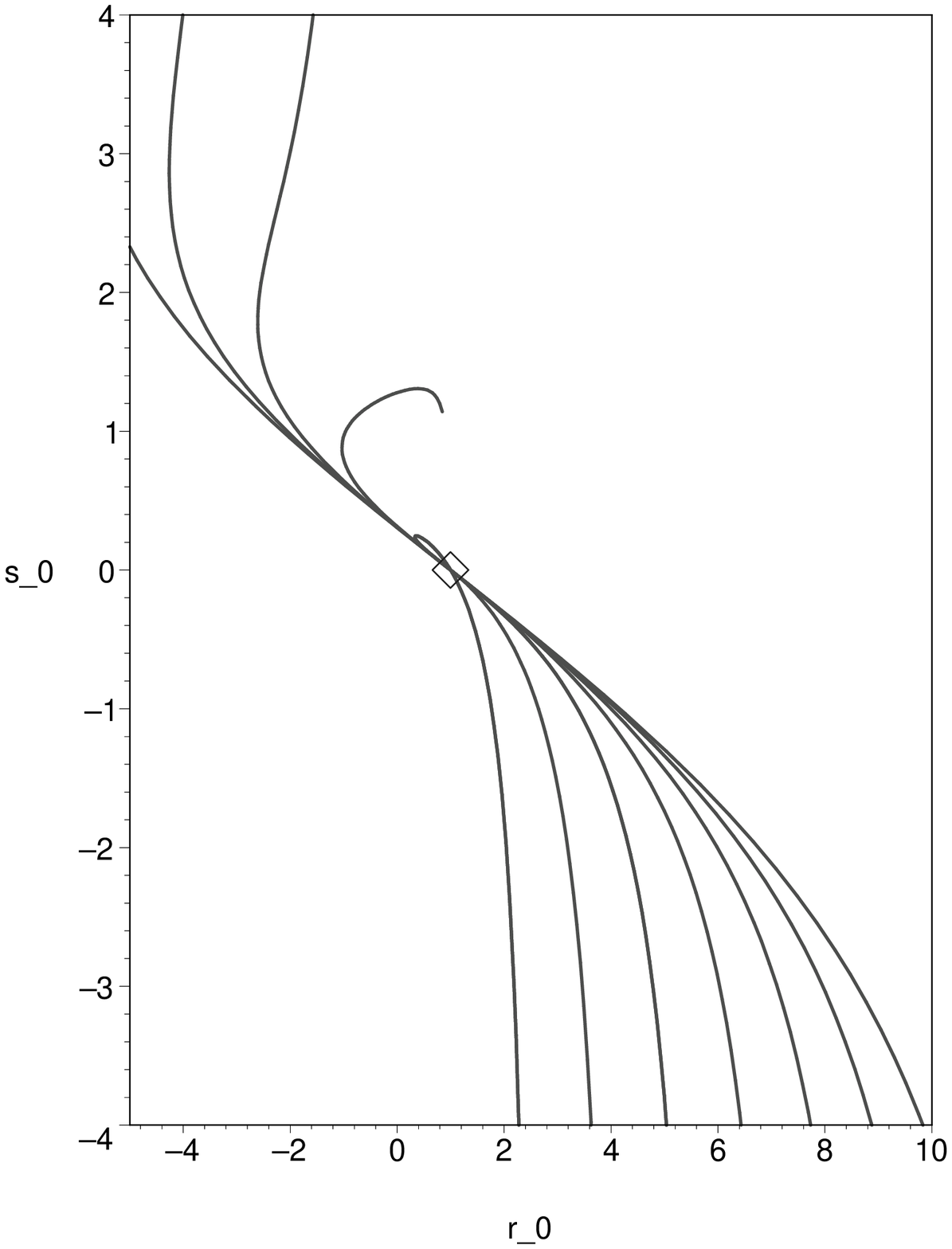}
\includegraphics[width=60mm,height=60mm]{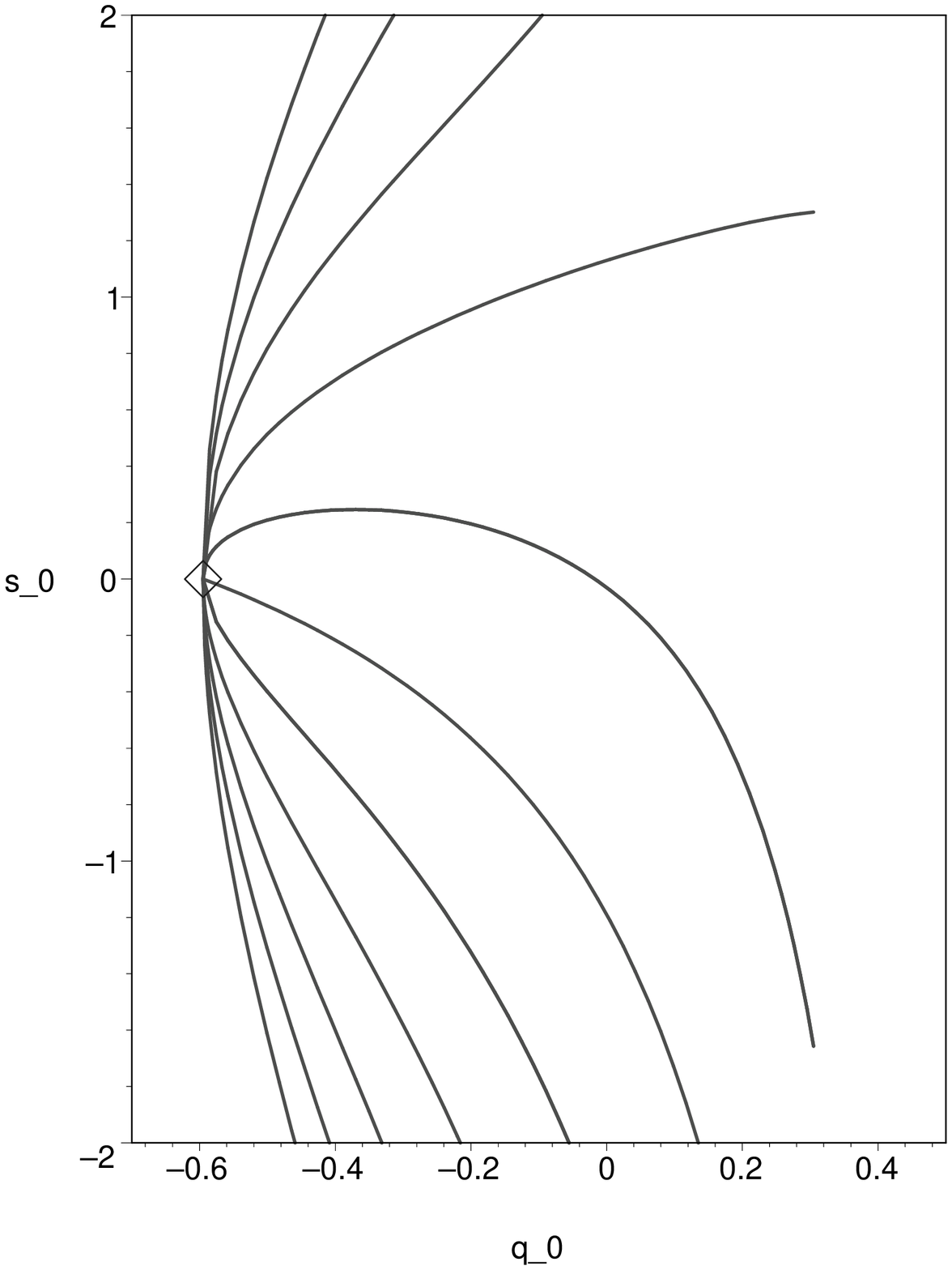}
  \end{center}
\caption{Present values of the statefinder parameters and the
  deceleration parameter for models with matter and quintessence with
  an exponential potential. The diamond represents the $\Lambda \rm CDM$
  model. \newline 
Top panel: From left to right the different curves have
$\lambda=-5,-4,-3,-2,-1,0,1,2,3,4,5$. \newline
Bottom panel:
From top to bottom the different curves have
  $\lambda=-5,-4,-3,-2,-1,0,1,2,3,4,5$. }
 \label{s0}
\end{figure}

For models with matter+quintessence+curvature, the Friedmann and 
energy conservation equations give  
\begin{eqnarray}
\dot H&=&-3H^2+\frac{1}{2M^2}\left( \frac{1}{2}\rho_{\rm m}-V(\phi) +\frac{2}{3}\rho_{\rm k}  \right) \\
\frac{1}{2}\dot\phi^2&=&3H^2M^2- \rho_{\rm m} -V(\phi)-\rho_{\rm k} \\
\dot\rho_{\rm m}&=&-3H\rho_{\rm m} \\
\dot \rho_{\rm k}&=&-2H\rho_{\rm k}, 
\end{eqnarray}
and 
\begin{eqnarray}
q&=&\frac{1}{2}\Omega_{\rm m}+2\Omega_{\rm kin}-\Omega_{\rm pot} \\
r&=&\Omega_{\rm m}+10\Omega_{\rm kin}+\Omega_{\rm pot} +3\sqrt{6\Omega_{\rm kin}}\frac{MV'}{\rho_{\rm c}}.   
\end{eqnarray}
As customary when discussing quintessence, we have introduced the 
Planck mass $M^2 = 1/8\pi G$.  Furthermore, we have defined   
$\Omega_{\rm kin}= \dot{\phi}^2 / 2\rho_{\rm c}$, and 
$\Omega_{\rm pot} = V(\phi) / \rho_{\rm c}$.
For an exponential potential, $V(\phi)=A\exp(-\lambda\phi/M)$, and looking 
at values at the present epoch, one gets 
\begin{eqnarray}
q_0&=&\frac{1}{2}\Omega_{\rm m0}+2\Omega_{\rm kin0}-\Omega_{\rm pot0} \\
r_0&=&\Omega_{\rm m0}+10\Omega_{\rm kin0}+\Omega_{\rm pot0}
-3\lambda\sqrt{6\Omega_{\rm kin0}}\Omega_{\rm pot0}. 
\end{eqnarray}
Eliminating $\Omega_{\rm pot0}$, using 
$\Omega_{\rm m0}+\Omega_{\rm kin 0}+ \Omega_{\rm pot 0}+ \Omega_{k 0}=1$, 
one obtains  
\begin{eqnarray}\label{q0}
q_0&=&\frac{3}{2}\Omega_{\rm m0} -(1-\Omega_{\rm k0})  +3\Omega_{\rm kin0} \\
\label{r0}
r_0&=&(1-\Omega_{\rm k0})+9\Omega_{\rm kin0} \nonumber \\ 
&-&3\lambda\sqrt{6\Omega_{\rm kin0}}(1-\Omega_{\rm k0}-\Omega_{\rm m0} -
\Omega_{\rm kin0}) 
\end{eqnarray}
By choosing for instance $\Omega_{\rm m0}=0.27$ and $\Omega_{\rm k0}=0$ we can plot
the values of $q_0$ and $r_0$ for varying $\Omega_{\rm kin0}$; see
figure \ref{r0q0}. 
As we can see from equations (\ref{q0})-(\ref{r0}), when  $\Omega_{\rm kin0}=0$,  $q_0$ and $r_0$
are independent of $\lambda$,  and have the same values 
as in the $\Lambda$CDM model. This is obvious, since taking 
away the kinetic 
term will reduce quintessence to LIVE. However,
when $\Omega_{\rm kin0}$ is slightly greater then $0$ we can make  $r_0$ as
large or as small as we like, by choosing $|\lambda|$ sufficiently large. 
There is no reason all quintessence models  should lie inside the 
constant-$w$-curve.
However, in order to get an accelerating universe today we must have
$\lambda^2<2$. But also for $\lambda^2<2$ the present values of
$q_0$ and $r_0$ can lie outside the constant-$w$-curve.
In fact, when we move on to the time-evolving statefinders, plotting
$q$ and $r$ as functions of time for given initial conditions, we obtain 
plots like figure \ref{rqexp}.
Here we have chosen as initial conditions $\Omega_{\rm m0}=0.27$ and
$\Omega_{\rm k0}=0$ as above,  and $h=0.71$. The
last initial condition, for the quintessence field, we have chosen to be 
$\phi_0=M/100$ combined with 
the overall constant $A$ in the potential chosen to give
$\Omega_{\rm kin0}$ as stated in the caption of figure \ref{rqexp}. This corresponds to the universe being matter dominated at earlier times.  
When $\Omega_{\rm pot0}\gg \Omega_{\rm kin0}$ we have high 
acceleration today. 
Choosing  $\Omega_{\rm kin0}=0$ will again give us $\Lambda$CDM.
The  three rightmost curves in the figure have 
$\lambda^2>2$ and no eternal 
acceleration, although the $\lambda=1.6$ universe accelerates
today. It seems that in order to get a universe close to what 
we observe, $r$ and $q$ for models with matter+quintessence
  with an exponential potential will essentially lie within the same area as
  matter+quiessence models. 
In figure \ref{s0} we have plotted the trajectories in the
$s_0$--$r_0$-plane and the $s_0$--$q_0$-plane for the same models as
in figure 
\ref{r0q0}, to be compared with figures 5c and 5d in Alam 
et al. (2003).  

\begin{figure}[htbp!]  
  \begin{center}
\includegraphics[width=60mm,height=60mm]{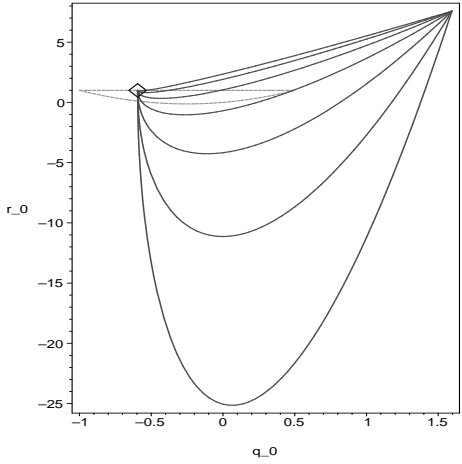}
  \end{center}
\caption{Present values of $q$ and $r$ for matter+quintessence with a
  power-law potential with $\alpha=2$. 
From top to bottom the different curves have
  $\phi_0=8M,4M,2M,M,\frac{M}{2},\frac{M}{4},\frac{M}{8}$.The diamond
  represents the $\Lambda \rm CDM$ model. The
  dotted curve shows  the area all matter+quiessence models must lie
  within at all times. }
 \label{pot-q0-r0}
\end{figure}

\begin{figure}[htbp!]  
  \begin{center}
\includegraphics[width=60mm,height=60mm]{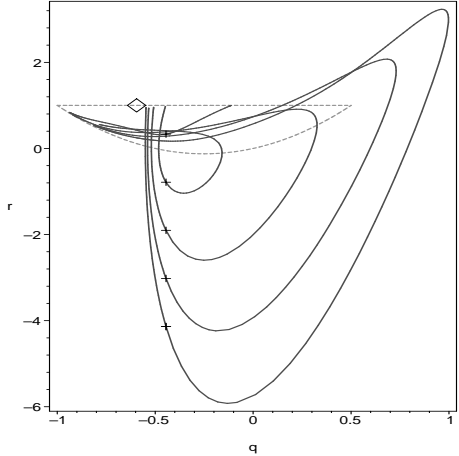}
  \end{center}
\caption{Time-evolution of $q$ and $r$ for models with matter and quintessence
  with a power-law potential. The crosses mark the present epoch,
  the diamond represents the present $\Lambda \rm CDM$ model.
 All models start out from the horizontal $\Lambda \rm CDM$ line and
  will eventually end up as a deSitter Universe ($q=-1,r=1$).  
The curve going deepest down has
  $\alpha=5$ and moving upwards we have $\alpha=4,3,2,1$.
The dotted curve
  shows the area all matter+quiessence models must lie within at all
  times. Obviously the same is not the case for matter+quintessence
  models.
}
 \label{rqpotfig}
\end{figure}

Choosing instead a power-law potential $V(\phi)=A\phi^{-\alpha}$ gives 
$V'=-\frac{\alpha}{\phi}V$
and
\begin{eqnarray}
q&=&\frac{1}{2}\Omega_{\rm m}+2\Omega_{\rm kin}-\Omega_{\rm pot} \\
r&=&\Omega_{\rm m}+10\Omega_{\rm kin}+\Omega_{\rm pot}
-3\alpha\frac{M}{\phi}\sqrt{6\Omega_{\rm kin}}\Omega_{\rm pot}.   
\end{eqnarray}
We see that for $\phi_0=M$ we get the same curves in the $q_0$--$r_0$-plane 
when varying
$\alpha$ as we got when varying $\lambda$ in the exponential potential, 
see figure \ref{r0q0}.  
We also see that varying $\phi_0$ for a given value of $\alpha$ is
essentially the same as varying $\alpha$. Figure \ref{pot-q0-r0} shows the
$q_0$--$r_0$-plane for the case $\alpha=2$.
Figure \ref{rqpotfig} shows an example of time-evolving
statefinders ($\phi_0=M$, $\Omega_{\rm kin0}=0.05$, $\Omega_{\rm m0}=0.27$
$\Omega_{\rm k0}=0$, $h$=0.71).  If one compares this plot with
figure 1b in \cite{alam}, the two do not quite agree. 
\cite{alam} do not give detailed information about the 
initial conditions for the quintessence field. Our initial conditions
correspond to a universe which was matter-dominated up to now, when
quintessence is taking over. 

\subsection{Dark energy fluid models}

We will now find expressions for $r$ and $s$ which are valid even if 
the dark energy does not have an equation of state of the form 
$p_{\rm x} = w \rho_{\rm x}$.  This is the case e.g. in the 
Chaplygin gas models (Kamenshchik, Moschella \& Pasquier 2001; Bilic, Tupper \& Viollier 2002).  The expression 
for the deceleration parameter can be written as 
\begin{equation}
q = \frac{1}{2}\left(1+3\frac{p_{\rm x}}{\rho_{\rm x}}\right)\Omega,
\label{eq:eq1.34}
\end{equation}
and using this in equation (\ref{eq:eq1.16}) we find 
\begin{eqnarray}
r&=&\left(1-\frac{3}{2}\frac{\dot{p}_{\rm x}}{H\rho_{\rm x}}\right)
\Omega \label{eq:eq1.35} \\ 
s & = & -\frac{1}{3H}\frac{\dot{p}_{\rm x}}{p_{\rm x}}. \label{eq:eq1.36}
\end{eqnarray}
If the universe contains only dark energy with an equation of state 
$p=p(\rho)$, then 
\begin{equation}
\dot{p} = \dot{\rho}\frac{\partial p}{\partial \rho} 
= -3H(\rho+p)\frac{\partial p}{\partial \rho},
\label{eq:eq1.37}
\end{equation}
which leads to 
\begin{eqnarray}
r & = & \left[1+\frac{9}{2}\left(1+\frac{p}{\rho}\right)\frac{\partial p}
{\partial \rho}\right]\Omega \label{eq:eq1.38} \\ 
s & = & \left(1+\frac{\rho}{p}\right)\frac{\partial p}{\partial \rho}. 
\label{eq:eq1.39}
\end{eqnarray}
If the universe contains cold matter and dark energy these expressions 
are generalized to 
\begin{eqnarray}
r&=& \left(1+\frac{9}{2}\frac{\rho_{\rm x}+p_{\rm x}}{\rho_{\rm m}
+\rho_{\rm x}}\frac{\partial p_{\rm x}}{\partial \rho_{\rm x}}\right)
\Omega \label{eq:eq1.40} \\ 
s & = & \left(1+\frac{\rho_{\rm x}}{p_{\rm x}}\right)
\frac{\partial p_{\rm x}}{\partial \rho_{\rm x}}. \label{eq:eq1.41}
\end{eqnarray}
The Generalized Chaplygin Gas (GCG) has an equation of state 
(Bento, Bertolami \& Sen 2002) 
\begin{equation}
p = -\frac{A}{\rho^\alpha},
\label{eq:eq1.42}
\end{equation}
and integration of the energy conservation equation gives 
\begin{equation}
\rho = \left[A+Ba^{-3(1+\alpha)}\right]^{\frac{1}{1+\alpha}}, 
\label{eq:eq1.43}
\end{equation}
where $B$ is a constant of integration.  This can be rewritten as 
\begin{equation}
\rho = \rho_0\left[A_{\rm s}+(1-A_{\rm s})x^{3(1+\alpha)}\right]^{\frac{1}{1+\alpha}},
\label{eq:eq1.44}
\end{equation}
where $\rho_0 = (A+B)^{1/(1+\alpha)}$, and $A_{\rm s}=A/(A+B)$.  For 
a flat universe with matter and a GCG, the Hubble parameter is given by 
\begin{equation}
\frac{H^2(x)}{H_0}=\Omega_{\rm m 0}x^3 + (1-\Omega_{\rm m 0}) 
\left[A_{\rm s}+(1-A_{\rm s})x^{3(1+\alpha)}\right]^{\frac{1}{1+\alpha}}.
\label{eq:eq1.45}
\end{equation}
This leads to the following expressions for $q(x)$ and $r(x)$: 
\begin{eqnarray}
q(x) &=& \frac{3}{2} \frac{\Omega_{\rm m 0} x^{3} + (1 - \Omega_{\rm m0})(1
- A_{\rm s}) \ v^{\frac{3}{\beta} - 1} \ x^{\beta}}{\Omega_{\rm m0} x^3 + (1 -
\Omega_{m0}) \ v^{3/{\beta}} } - 1 \label{eq:eq1.46} \\ 
r(x) &=& 1 - 3 \ \frac{x}{h^2(x)}f(x) + \frac{3}{2} \ \frac{x^2}{h^2(x)}
f'(x)  \label{eq:eq1.47}, 
\end{eqnarray}
where $\beta=3(1+\alpha)$, $h(x)=H(x)/H_0$, and 
\begin{eqnarray}
v & = & A_{\rm s} + (1 - A_{\rm s})x^{\beta} \label{eq:eq1.48} \\
f(x) & = & \Omega_{\rm m0} x^{2} + (1 - \Omega_{\rm m 0})(1- A_{\rm s}) 
v^{\frac{3}{\beta} - 1} \ x^{\beta-1}. \label{eq:eq1.49}
\end{eqnarray}
In the $r$-$s$ plane, the GCG models will lie on curves given by 
(see Gorini, Kamenshchik \& Moschella 2003)
\begin{equation}
r = 1-\frac{9}{2}\frac{s(s+1)}{\alpha}.
\end{equation}
We note that a recent comparison of GCG models with SNIa data found 
evidence for $\alpha > 1$ (Bertolami et al. 2004).  

\subsection{Cardassian models}

As an alternative to adding a negative-pressure component to the 
energy-momentum tensor of the Universe, one can take the view that 
the present phase of accelerated expansion is caused by gravity 
being modified, e.g. by the presence of large 
extra dimensions.  For a general discussion of extra-dimensional 
models and statefinder parameters, see Alam \& Sahni (2002). 

As an example, we will consider the Modified Polytropic  
Cardassian ansatz (MPC) (Freese \& Lewis 2002; Gondolo \& Freese 2003), 
where the Hubble parameter is given by 
\begin{equation}
H(x) = H_0 \sqrt{\Omega_{\rm m 0} x^3  \Bigl( 1 + u \Bigr)^{1/\psi} }, 
\label{eq:eq1.50}
\end{equation}
with 
\begin{equation}
u = u(x) = \frac{\Omega_{\rm m 0}^{-\psi} \ - \ 1}{x^{3(1-n)\psi}},  
\label{eq:eq1.51}
\end{equation}
and where $n$ and $\psi$ are new parameters ($\psi$ is usually called $q$ in the 
literature, but we use a different notation here to avoid confusion with 
the deceleration parameter).  For this model, the 
deceleration parameter is given by 
\begin{equation}
q(x) = \frac{3}{2} \Biggl[ \frac{1 + nu}{1 + u} \Biggr] - 1
\label{eq:eq1.52}
\end{equation}
and the statefinder $r$ by 
\begin{eqnarray}
r(x) &=& 1 - \frac{9}{4} \  \frac{1 + nu}{1 + u} \Biggl[ 1 + \frac{u 
(1-n) - (1 + nu)}{1 + u} \nonumber \\ 
&-& 2q \frac{(1 - n)^2 u}{(1 + u)(1 + nu)}.  \Biggr] \label{eq:eq1.53}
\end{eqnarray}

\subsection{The luminosity distance to third order in $z$} 

The statefinder parameters appear when one expands the luminosity 
distance to third order in the redshift $z$.  
Although this expansion has been presented earlier 
(Chiba \& Nakamura 1998; Visser 2003) in a slightly different notation,  
we carry out this derivation here for completeness.  The luminosity 
distance is given by the expression 
\begin{equation}
d_{\rm L} = \frac{1+z}{H_0\sqrt{|\Omega_{\rm k 0}}}{\cal S}_k(\sqrt{|\Omega_{\rm k0}|}
I),
\label{eq:eq1.54}
\end{equation}
where ${\cal S}_{\rm k}(x) = \sin x$ for $k=1$, ${\cal S}_{\rm k}(x)= x$ for $k=0$,  
${\cal S}_{\rm k}(x) = \sinh x $ for $k=-1$, and 
\begin{equation}
I = H_0 \int_0 ^ z \frac{dz}{H(z)}.
\label{eq:eq1.55}
\end{equation}
Using the approximation 
\begin{equation}
{\cal S}_{\rm k}(x) \approx x - \frac{k}{6}x^3,
\label{eq:eq1.56}
\end{equation}
one finds to third order in $z$
\begin{equation}
H_0d_{\rm L} \approx (1+z)I\left(1+\frac{1}{6}\Omega_{\rm k0}I^2\right).
\label{eq:eq1.57}
\end{equation}
Taylor expanding the Hubble parameter to second order in $z$, we have 
\begin{equation}
I \approx \int_0 ^ z \frac{dz}{1+\gamma z+\kappa z^2},
\label{eq:1.58}
\end{equation}
with 
\begin{eqnarray}
\gamma & = & \frac{1}{H_0}\left(\frac{dH}{dz}\right)_{z=0}, \label{eq:1.59} \\ 
\kappa & = & \frac{1}{2H_0}\left(\frac{d^2 H_0}{dz^ 2}\right)_{z=0}. 
\label{eq:eq1.60} 
\end{eqnarray}
By using $(1+y)^{-1}\approx 1-y+y^2$ with $y=\gamma z+\kappa z^2$, 
we get to third order in $z$ 
\begin{eqnarray}
(1+z)I &\approx& (1+z)\int_0^z\left[1-\gamma z+(\gamma^2 - \kappa)z^ 2\right]dz 
\nonumber \\
&= & z + \left(1-\frac{1}{2}\gamma\right)z^2 \nonumber \\ 
&-&\left(\frac{1}{2}\gamma+
\frac{1}{3}\kappa - \frac{1}{3}\gamma^2\right)z^3. \label{eq:eq1.61}
\end{eqnarray}
We wish to express $\gamma$ and $\kappa$ in terms of the deceleration 
parameter 
\begin{equation}
q_0 = - \left(\frac{\ddot{a}a}{\dot{a}^2}\right)_{t=t_0},
\label{eq:eq1.62}
\end{equation}
and the statefinder 
\begin{equation}
r_0 = \left(\frac{a^2 \stackrel{{\bf\ldots}}{a}}{\dot{a}^3}\right)_{t=t_0}.
\label{eq:eq1.63}
\end{equation}
From $H = \dot{a} / a$ one finds 
\begin{eqnarray}
\dot{H} & = & -(1+q)H^2, \label{eq:eq1.64} \\ 
\ddot{H} &= & (r+3q+2)H^3. \label{eq:eq1.65} 
\end{eqnarray}
From $a=(1+z)^{-1}$, one gets $H=-\dot{z}/(1+z)$, and hence 
\begin{eqnarray}
\frac{d}{dz} & = & -\frac{1}{(1+z)H}\frac{d}{dt}, \label{eq:eq1.66} \\ 
\frac{d^2}{dz^2} & = & \frac{1}{(1+z)H}\frac{d}{dt}\left[\frac{1}{(1+z)H}
\frac{d}{dt}\right]. \label{eq:eq1.67} 
\end{eqnarray}
After some algebra one then finds 
\begin{eqnarray}
\gamma & = & 1+q_0, \label{eq:eq1.68} \\ 
\kappa & = & \frac{1}{2}(r_0 - q_0^2). \label{eq:eq1.69}
\end{eqnarray}
Substitution of these expressions for $\gamma$ and $\kappa$ in 
(\ref{eq:eq1.61}) gives
\begin{eqnarray}
(1+z)I &\approx& z\Biggl[1+\frac{1}{2}(1-q_0)z-\frac{1}{6}(1+r_0 \nonumber \\ 
&-&q_0-3q_0^2)
z^2\Biggr],
\label{eq:eq1.70}
\end{eqnarray}
and equation (\ref{eq:eq1.57}) then finally leads to 
\begin{eqnarray} 
d_{\rm L} &\approx& \frac{z}{H_0}\Biggl[1+\frac{1}{2}(1-q_0)z 
-\frac{1}{6}(1+r_0-q_0\nonumber \\ 
&-&3q_0^2-\Omega_{\rm k 0})z^2\Biggr].
\label{eq:eq1.71}
\end{eqnarray}

One can also find an expression for the present value of the time 
derivative of the equation of state parameter $w$ in terms of the 
statefinder $r_0$.  A Taylor expansion to first order in $z$ gives 
\begin{equation}
w(z) = w_0 - \frac{\dot{w}(t_0)}{H_0}z, 
\label{eq:eq1.72}
\end{equation}
From equation (\ref{eq:eq1.15}) we have 
\begin{equation}
w_0 = \frac{2q_0-\Omega_{\rm m 0}-\Omega_{\rm x 0}}{3\Omega_{\rm x 0}},
\label{eq:eq1.73}
\end{equation}
and from equation (\ref{eq:eq1.25}) we get 
\begin{equation}
\dot{w}(t_0) = \frac{2}{3}\frac{H_0}{\Omega_{\rm x 0}}
\left\{ \Omega_{\rm m 0}+\left[1+\frac{9}{2}w_0(1+w_0)\right]
\Omega_{\rm x 0} -r_0 \right \}. 
\label{eq:eq1.74}
\end{equation}
Hence 
\begin{equation}
w(z) \approx w_0 - \frac{2}{3}\left[1+\frac{9}{2}w_0(1+w_0)
+\frac{\Omega_{\rm m 0}-r_0}{\Omega_{\rm x 0}}\right]z.
\label{eq:eq1.75}
\end{equation}

\section{Lessons drawn from current SNIa data}

In this section we will consider the SNIa data presently available, 
in particular whether one can use them to learn about the statefinder 
parameters.  We will use the recent collection of SNIa data in 
Riess et al. (2004), their `gold' sample consisting of 157 supernovae 
at redshifts between $\sim 0.01$ and $\sim 1.7$.    

\subsection{Model-independent constraints}

\begin{figure}[ht]
\begin{center}
\includegraphics[width=60mm,height=60mm]{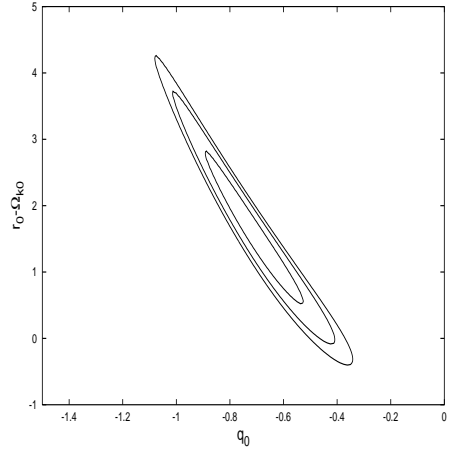}
\caption{Likelihood contours (68, 95 and 99\%) resulting from a fit 
of the expansion of the luminosity distance to third order in $z$.}
\label{fig:qrcontour}
\end{center}
\end{figure}

The approximation to $d_{\rm L}$ in equation (\ref{eq:eq1.71}) is independent 
of cosmological model, the only assumption made is that the Universe 
is described by the Friedmann-Robertson-Walker metric.  We see that, 
in addition to $H_0$, this third-order expansion of $d_{\rm L}$ depends 
on $q_0$ and the combination $r_0-\Omega_{\rm k 0}$.  Fitting these 
parameters to the data, we find the constraints shown in 
Fig. \ref{fig:qrcontour}.  The results are consistent with those 
of similar analyses in Caldwell \& Kamionkowski (2004) and Gong (2004). 
\begin{figure}[ht]
\begin{center}
\includegraphics[width=60mm,height=60mm]{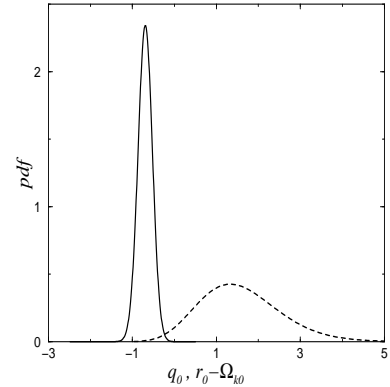}
\caption{Marginalized probability distributions for $q_0$ (full line) and 
$r_0 - \Omega_{\rm k 0}$ (dotted line).}
\label{fig:pdfs}
\end{center}
\end{figure}  
In figure \ref{fig:pdfs} we show the marginalized distributions for 
$q_0$ and $r_0-\Omega_{\rm k 0}$.    
We note that the supernova data firmly support an accelerating universe, 
$q_0 < 0$ at more than 99\% confidence.  However, about the statefinder 
parameter $r_0$, little can be learned without an external constraint on 
the curvature.  Imposing a flat universe, e.g. by inflationary 
prejudice or by invoking the  CMB peak positions, 
there is still a wide range of allowed values for $r_0$.  This is an 
indication of the limited ability of the current SNIa data to place 
constraints on models of dark energy.  There is only limited information 
on anything beyond the present value of the second derivative of the 
Hubble parameter.  

\begin{figure}[ht]
\begin{center}
\includegraphics[width=60mm,height=60mm]{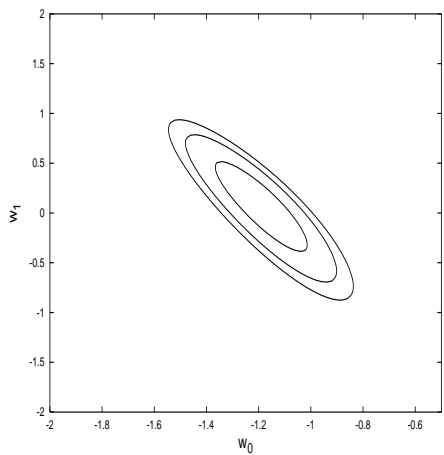}
\caption{Likelihood contours (68, 95 and 99\%) for the coefficients 
$w_0$ and $w_1$ in the linear approximation to the equation of state 
$w(z)$ of dark energy, resulting from a fit 
of the expansion of the luminosity distance to third order in $z$.}
\label{fig:w0w1contour}
\end{center}
\end{figure}

Under the assumption of a spatially flat universe, $\Omega_{\rm k 0}=0$, 
with $\Omega_{\rm m 0}=0.3$, one can use equations 
(\ref{eq:eq1.73}) and (\ref{eq:eq1.75}) to obtain constraints on $w_0$ and 
$w_1$ in the expansion $w(z) = w_0 + w_1 z$ of the equation of state of
dark energy.  The resulting likelihood contours are shown in 
figure \ref{fig:w0w1contour}.  As can be seen in this figure, there 
is no evidence for time evolution in the equation of state, the 
observations are consistent with $w_1 = 0$.  
The present supernova data show a slight preference for a dark energy 
component of the `phantom' type with $w_0 < -1$ (Caldwell 2002). 
Note, however, that 
the relatively tight contours obtained here are caused by the 
strong prior $\Omega_{\rm m0}=0.3$. It should also be noted that the 
third-order expansion of $d_{\rm L}$ is not a good approximation to the exact 
expression for high $z$ and in some regions of the parameter 
space.    

\subsection{Direct test of models against data}

The standard way of testing dark energy models against data is by 
maximum likelihood fitting of their parameters.  In this subsection 
we will consider the following models: 
\begin{enumerate}
\item The expansion of $d_{\rm L}$ to second order in $z$, with 
$h$ and $q_0$ as parameters.
\item The third-order expansion of $d_{\rm L}$, with $h$, $q_0$, 
and $r_0-\Omega_{\rm k 0}$ as parameters. 
\item Flat $\Lambda{\rm CDM}$ models, with $\Omega_{\rm m 0}$ and $h$ 
as parameters to be varied in the fit. 
\item $\Lambda{\rm CDM}$ with curvature, so that $\Omega_{\rm m 0 }$, 
$\Omega_{\Lambda 0}$ (the contribution of the cosmological constant to the 
energy density in units of the critical density, evaluated at the present 
epoch), and $h$ are varied in the fits. 
\item Flat quiessence models, that is, models with a constant equation 
of state $w$ for the dark energy component.  The parameters to be 
varied in the fit are $\Omega_{\rm m 0}$, $w$, and $h$. 
\item The Modified Polytropic Cardassian (MPC) ansatz, 
with $\Omega_{\rm m 0}$, 
$q$, $n$, and $h$ as parameters to be varied. 
\item The Generalized Chaplygin Gas (GCG), with $\Omega_{\rm m 0}$, $A_{\rm s}$, 
$\alpha$, and $h$ as parameters to be varied. 
\item The ansatz of Alam et al. (2003), 
\begin{equation}
H = H_0\sqrt{\Omega_{\rm m 0}x^3 + A_0 + A_1 x + A_2 x^2}, 
\label{eq:alamansatz}
\end{equation}
where we restrict ourselves to flat models, so that 
$A_0 = 1 - \Omega_{\rm m 0} -A_1 - A_2$.  The parameters to be 
varied are $\Omega_{\rm m 0}$, $A_1$, $A_2$, and $h$. 
\end{enumerate}
Note that these models have different numbers of free parameters.  
To get an idea of which of these models is 
actually preferred by the data, we therefore employ the Bayesian Information 
Criterion (BIC) (Schwarz 1978; Liddle 2004).  This is an approximation 
to the Bayes factor (Jeffreys 1961), which gives the posterior 
probability of one model relative to another assuming that there is 
no objective reason to prefer one of the models prior to fitting the data. 
It is given by 
\begin{equation}
{\cal B} = \chi^2 _{\rm min} + N_{\rm par} \ln N_{\rm data}, 
\label{eq:bic}
\end{equation}
where $\chi^2_{\rm min}$ is the minimum value of the $\chi^2$ for 
the given model against the data, $N_{\rm par}$ is 
the number of free parameters, 
and $N_{\rm data}$ is the number of data points used in the fit.  
As a result of the approximations made in deriving it, ${{\cal B}}$ is given 
in terms of the minimum $\chi^2$, even though it is related to the 
integrated likelihood.    The preferred model is the one which 
minimizes ${\cal B}$.  
In table \ref{tab:tab1} we have collected the results for the best-fitting 
models.  
\begin{table}
\begin{center}
\caption{Results of fitting the models considered in this subsection to 
the SNIa data}
\begin{tabular}{l|l|c|l} 
\hline 
Model & $\chi^2_{\rm min}$ &\# parameters & ${{\cal B}}$ \\ \hline \hline
2. order expansion & 177.1 & 2 &  187.2 \\ 
3. order expansion & 162.3 & 3 &  177.5 \\   
Flat $\Lambda{\rm CDM}$ & 163.8 & 2 & 173.9 \\  
$\Lambda{\rm CDM}$ with curvature& 161.2 & 3 & 176.4 \\ 
flat + constant EoS & 160.0 & 3 & 175.2 \\ 
MPC       & 160.3       &  4  &   180.5    \\ 
GCG       & 161.4 &  4 & 181.6 \\ 
Alam et al. & 160.5 & 4 & 180.7 \\ \hline \hline               
\end{tabular}
\label{tab:tab1}
\end{center}
\end{table} 
When comparing models using the BIC, the rule of thumb is that a difference 
of 2 in the BIC is positive evidence against the model with the larger value, 
whereas if the difference is 6 or more, the evidence against the model 
with the larger BIC is considered strong.    
The second-order expansion of $d_{\rm L}$ is then clearly disfavoured, thus 
the current supernova data give information, although limited, on 
$r_0 - \Omega_{\rm k 0}$.  We see that there is no indication in the 
data that curvature should be added to the $\Lambda{\rm CDM}$ model. 
Also, the last three models in table 
\ref{tab:tab1} seem to be disfavoured.   
One can conclude that there is no evidence in the current data that 
anything beyond flat $\Lambda{\rm CDM}$ is required.  This does not, 
of course, rule out any of the models, but tells us that the current 
data are not good enough to reveal physics beyond 
spatially flat $\Lambda{\rm CDM}$.  A similar conclusion was reached 
by Liddle (2004) using a more extensive collection of cosmological 
data sets and considering adding parameters to the flat $\Lambda{\rm CDM}$ 
model with scale-invariant adiabatic fluctuations.

\subsection{Statefinder parameters from current data}

If the luminosity distance $d_{\rm L}$ is found as a function of 
redshift from observations of standard candles, one can obtain 
the Hubble parameter formally from 
\begin{equation}
H(x) = \left[\frac{d}{dx}\left(\frac{d_{\rm L}}{x}\right)\right]^{-1}.
\label{eq:eq2.1}
\end{equation}
However, since  observations always contain noise, this relation 
cannot be applied straightforwardly to the data. Alam et al. (2003) 
suggested parametrizing the dark energy density as a second-order 
polynomial in $x$, $\rho_{\rm x}=\rho_{\rm c 0}(A_0+A_1x+A_2x^2)$, 
leading to a Hubble parameter of the form  
\begin{equation}
H(x) = H_0\sqrt{\Omega_{\rm m0}x^3 + A_0 + A_1 x + A_2 x^2}, 
\label{eq:eq2.2}
\end{equation}
and fitting $A_0$, $A_1$, and $A_2$ to data.  This parametrization 
reproduces exactly the cases  $w=-1$ ($A_1 = A_2 = 0$), 
$w=-2/3$ ($A_0 = A_2 = 0$), and $w=-1/3$ ($A_0=A_1=0$), 
and the luminosity distance-redshift relationship is given by 
\begin{equation}
d_{\rm L} = \frac{1+z}{H_0}\int _1 ^{1+z}\frac{dx}
{\sqrt{\Omega_{\rm m 0}x^3 + A_0 + A_1 x + A_2 x^2}}.
\label{eq:eq2.3}
\end{equation}
Having fitted the parameters $A_0$, $A_1$, and $A_2$ to e.g. supernova 
data using (\ref{eq:eq2.3}), one can then find $q$ and  $r$ 
by substituting equation (\ref{eq:eq2.2}) into (\ref{eq:eq1.9}) and 
(\ref{eq:eq1.14}): 
\begin{eqnarray}
q(x) & = & \frac{1}{2}\left(1-\frac{A_2x^2+2A_1x+3A_0}{\Omega_{\rm m 0} x^3 
+ A_2 x^2 + A_1 x + A_0}\right) \label{eq:eq2.4} \\ 
r(x) & = & \frac{\Omega_{\rm m 0}x^3 + A_0}{\Omega_{\rm m 0} x^3 
+ A_0 + A_1x + A_2 x^2}, \label{eq:eq2.5}
\end{eqnarray}
and furthermore the statefinder $s$ is found to be 
\begin{equation}
s(x) = \frac{2}{3}\frac{A_1 x + A_2 x^2}{3A_0+2A_1x+A_2x^2}, 
\label{eq:eq2.6}
\end{equation}
and the equation of state is given by 
\begin{equation}
w(x) = -1 + \frac{1}{3}\frac{A_1x+2A_2x^2}{A_0+A_1x+A_2x^2}.
\label{eq:eq2.65}
\end{equation}
The simulations of Alam et al. (2003) indicated that the statefinder 
parameters can be 
reconstructed well from simulated data based on a range of dark energy 
models, so we will for now proceed on the assumption that the parametrization 
in equation (\ref{eq:eq2.2}) is adequate for the purposes of extracting 
$q$, $r$ and $s$ from SNIa data.  We comment this issue in section 4. 

\begin{figure}[ht]
\begin{center}
\includegraphics[width=40mm,height=40mm]{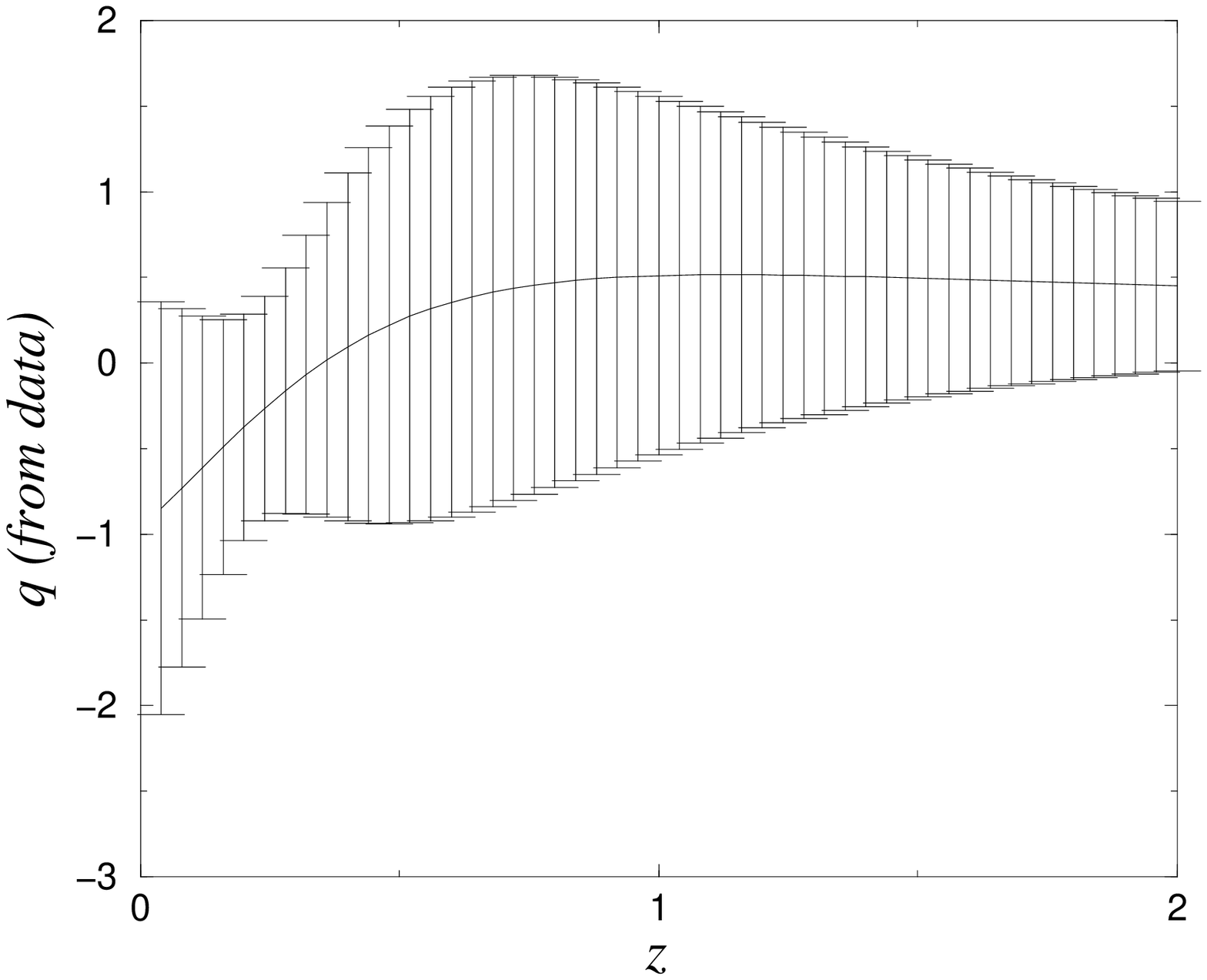}
\includegraphics[width=40mm,height=40mm]{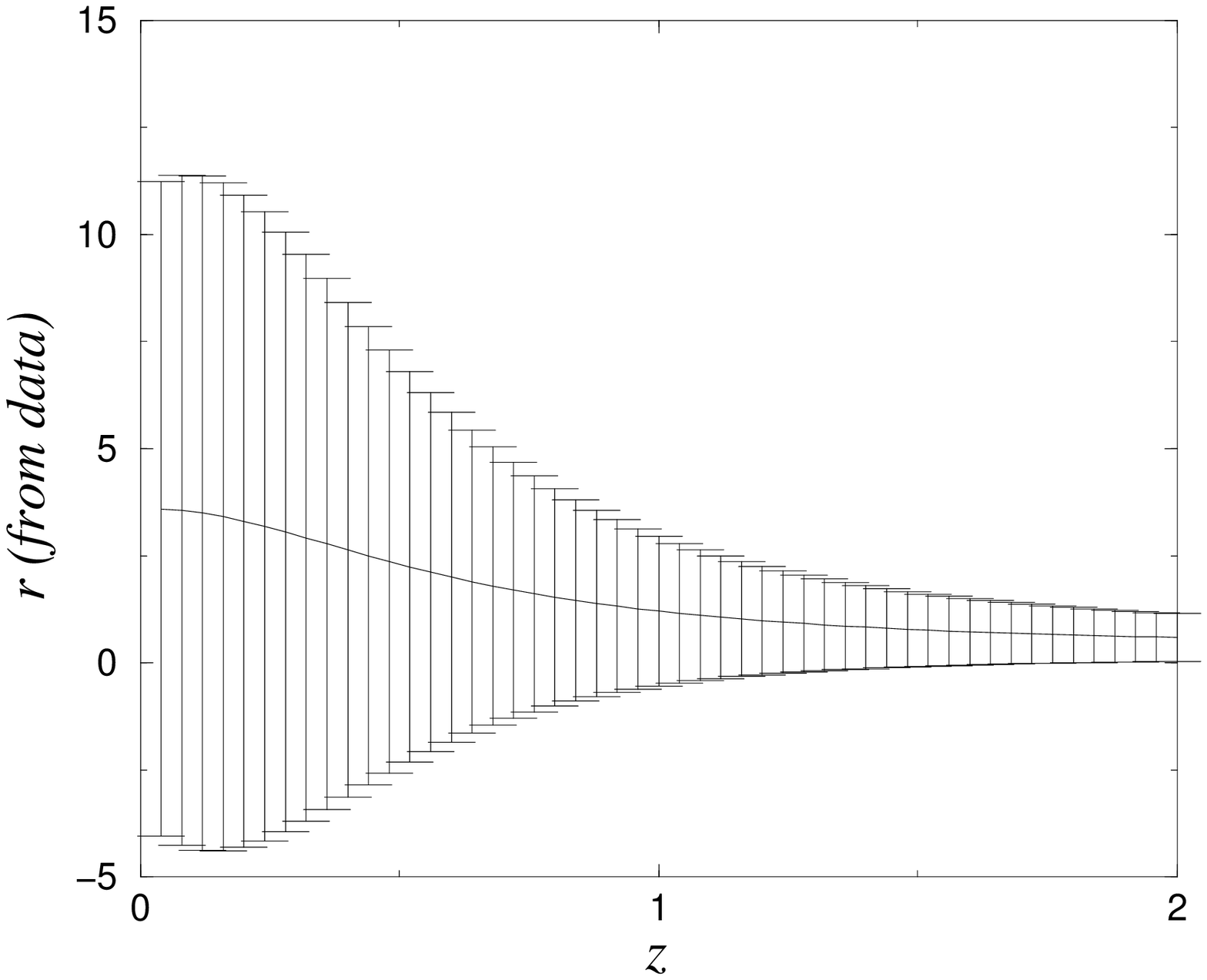}
\includegraphics[width=40mm,height=40mm]{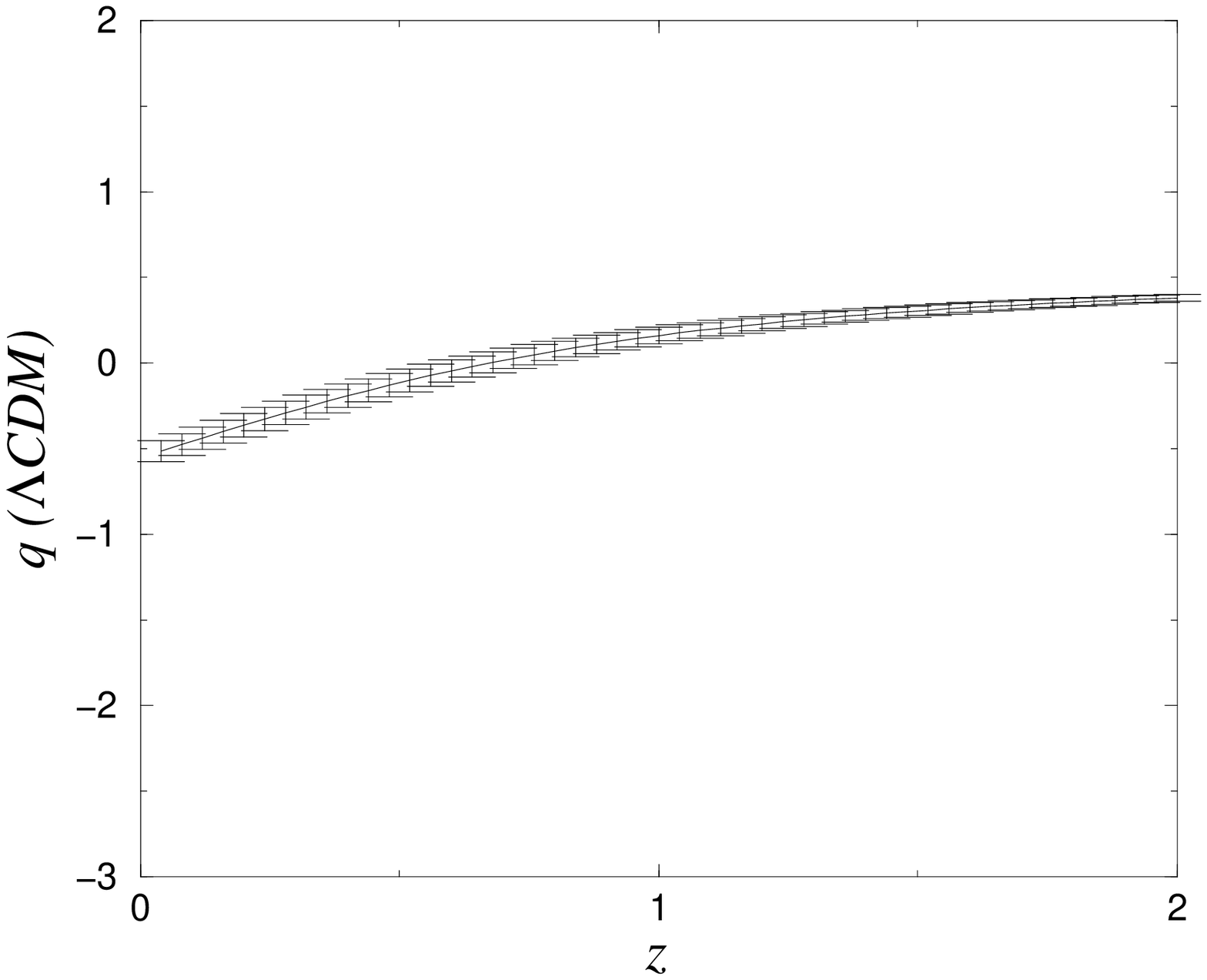}
\includegraphics[width=40mm,height=40mm]{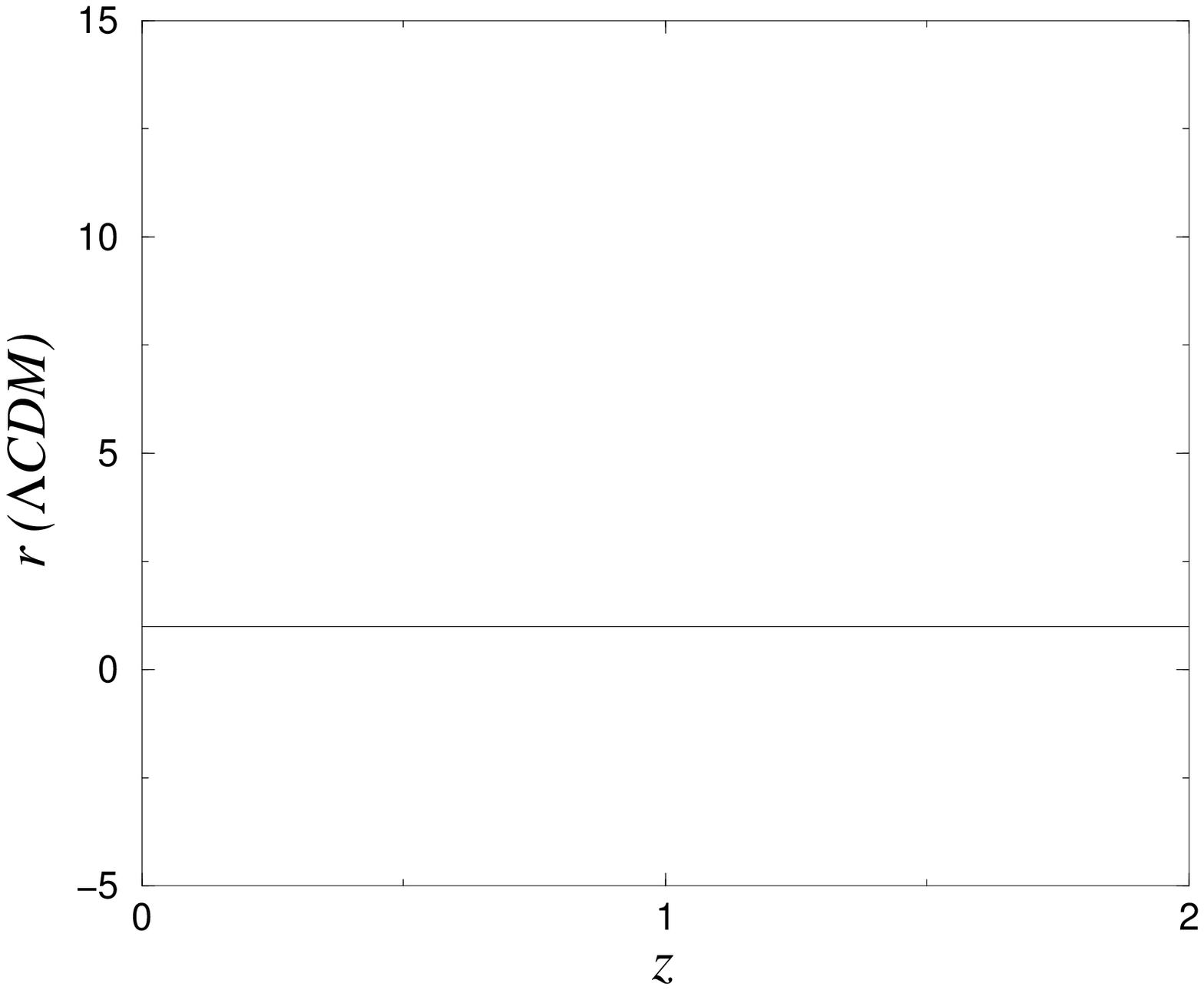}
\includegraphics[width=40mm,height=40mm]{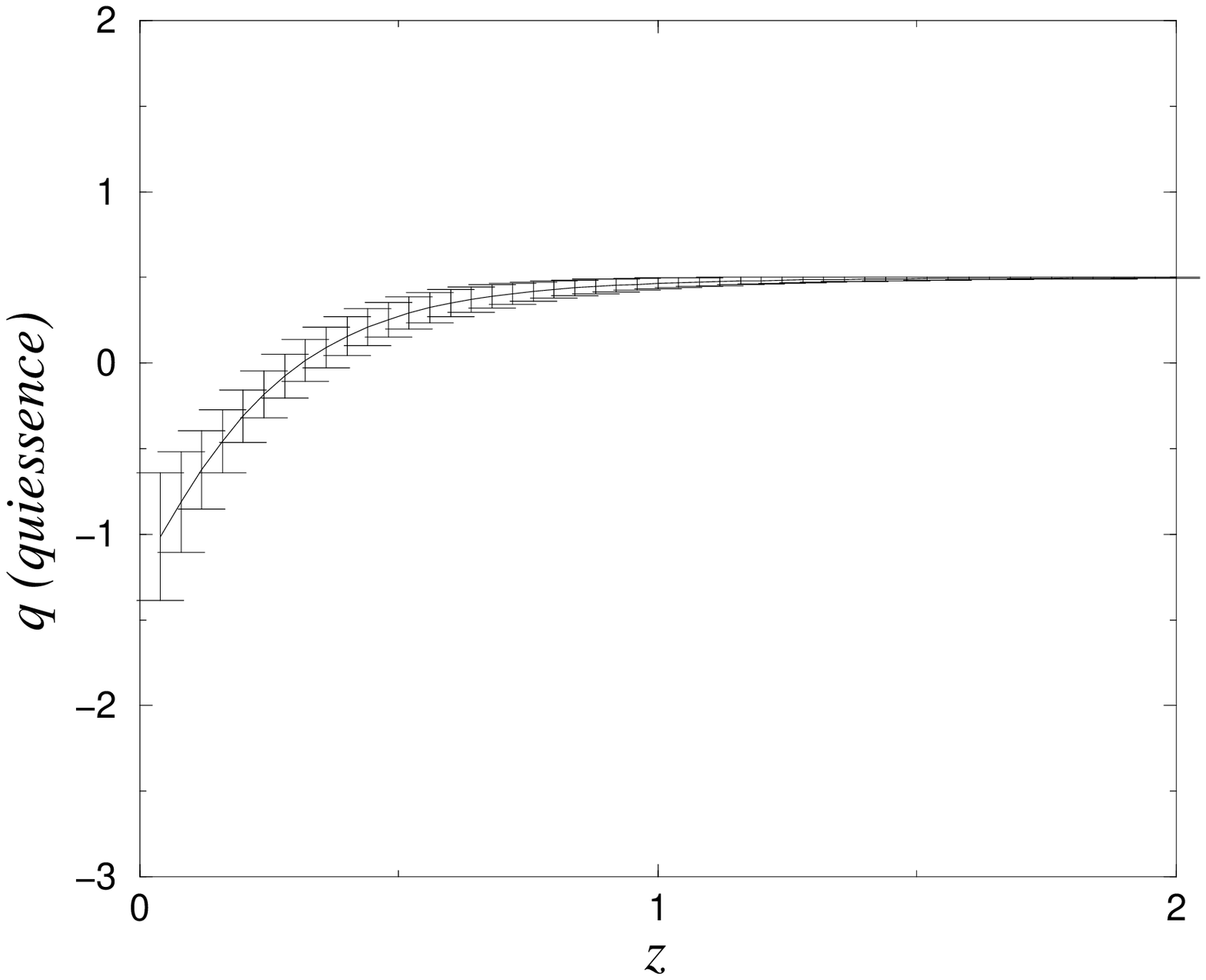}
\includegraphics[width=40mm,height=40mm]{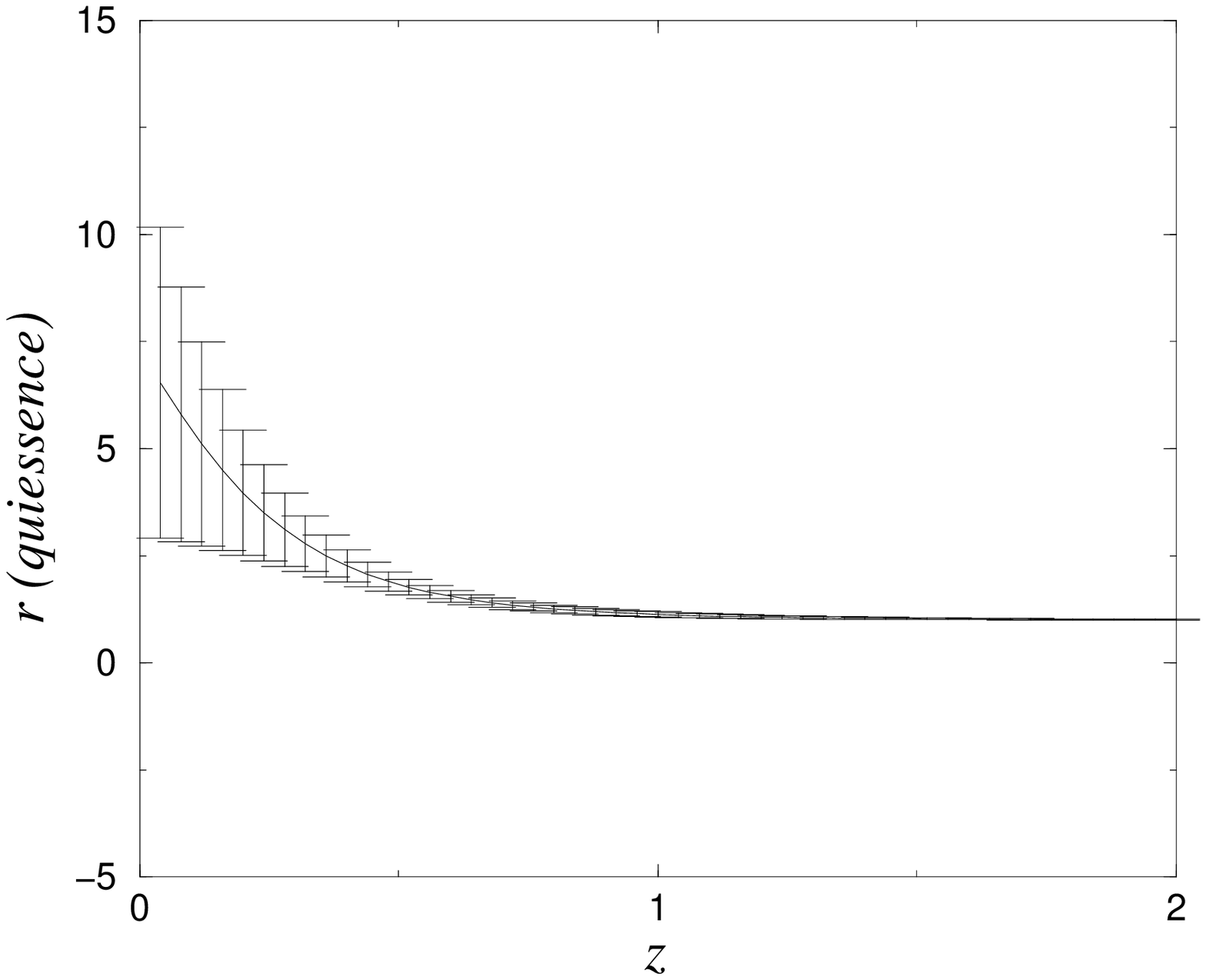}
\includegraphics[width=40mm,height=40mm]{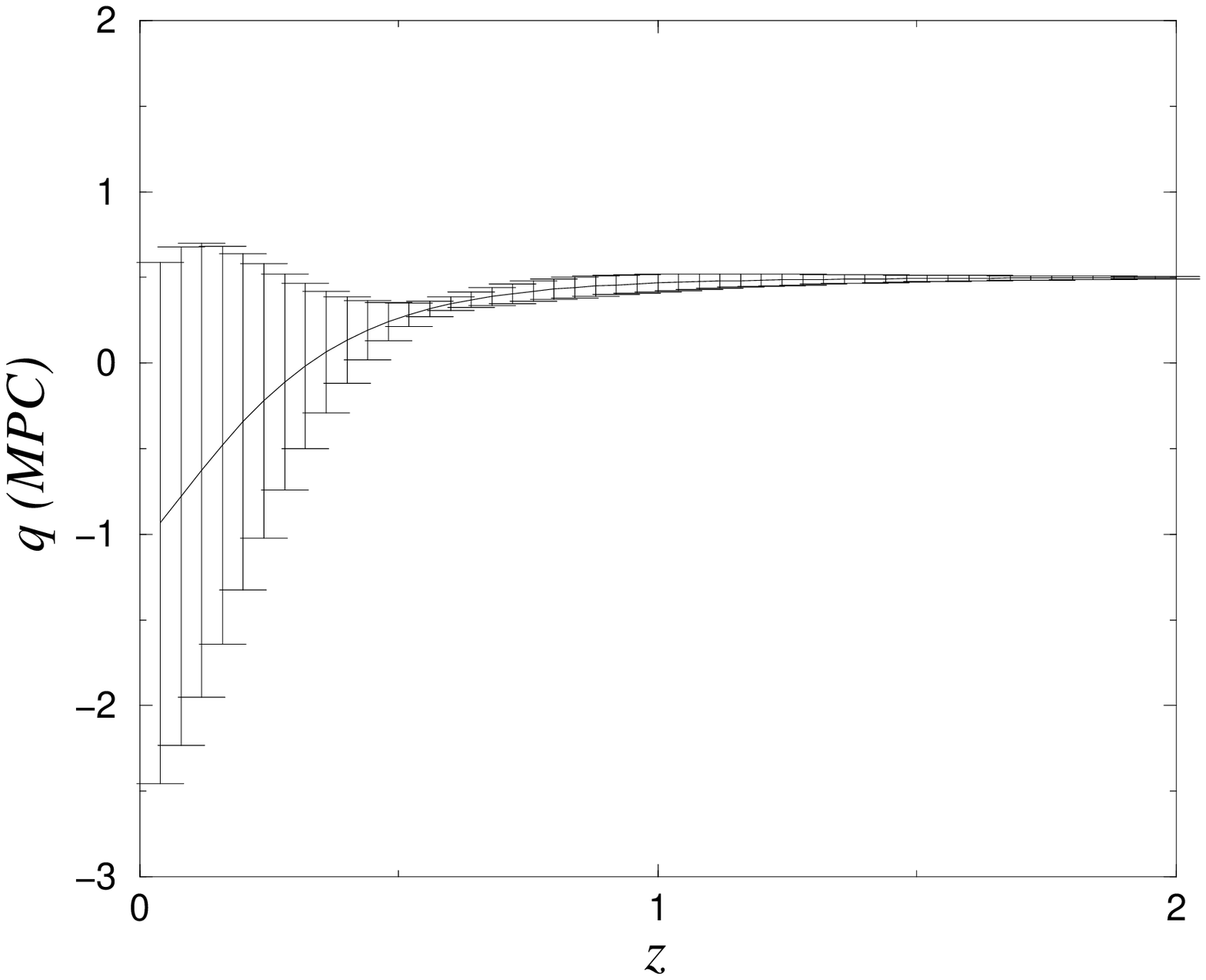}
\includegraphics[width=40mm,height=40mm]{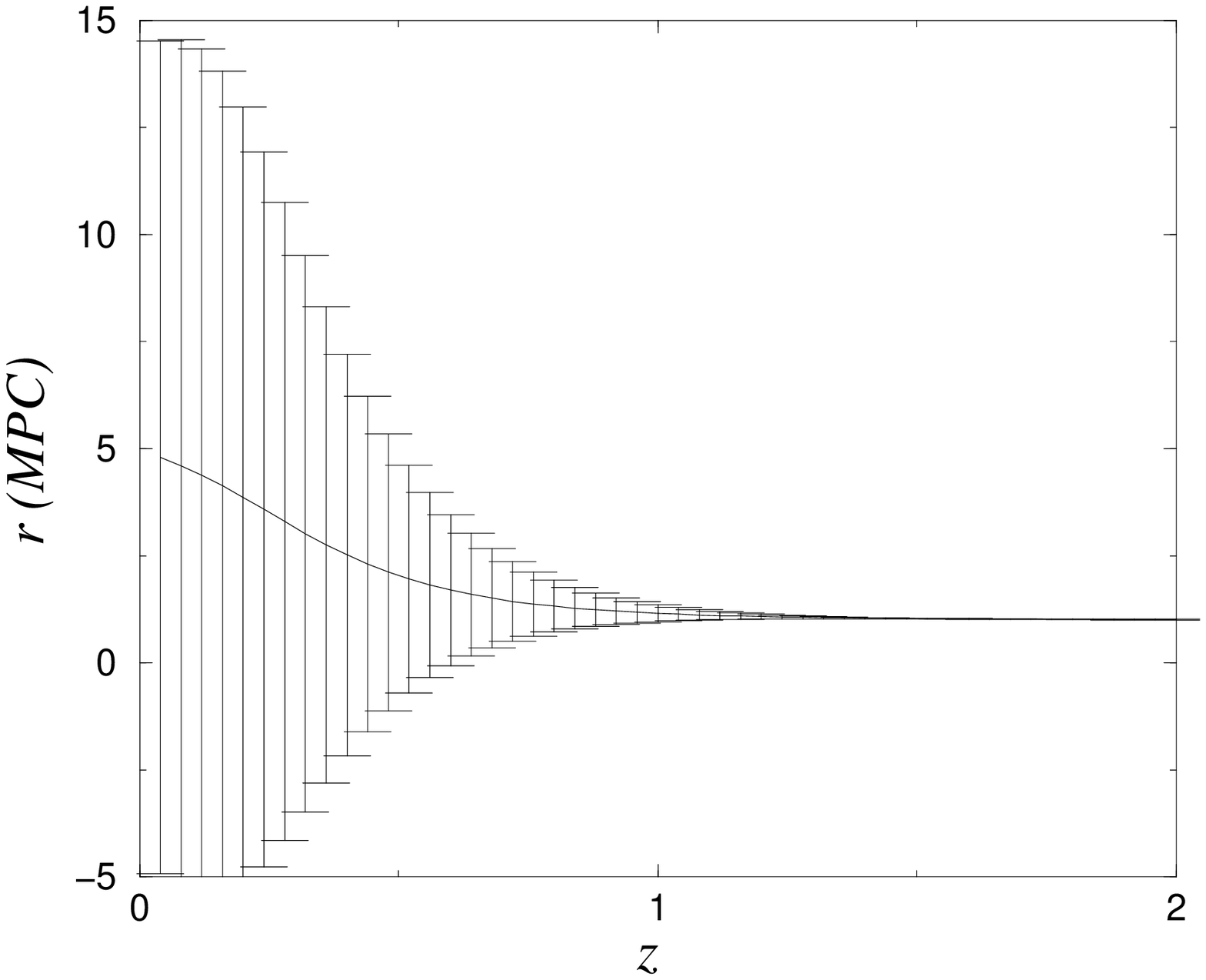}
\caption{The deceleration parameter $q$ and the statefinder 
$r$ 
extracted from current SNIa data using the Alam parametrization of 
$H$ (top row), for $\Lambda{\rm CDM}$ (second row), quiessence 
(third row), and the Modified Polytropic Cardassian ansatz 
(bottom row)}
\label{fig:qrfromdata}
\end{center}
\end{figure}

In figure \ref{fig:qrfromdata} we show the deceleration parameter 
$q$ and the statefinder $r$ extracted from the current SNIa data. 
The error bars in the figure are $1\sigma$ limits.  We have also plotted 
the model predictions for the same quantities (based on best-fitting 
parameters with errors) for $\Lambda{\rm CDM}$, quiessence, and the 
MPC.  The figure shows that all models are consistent at the $1\sigma$ level 
with $q$ and $r$ extracted using equation (\ref{eq:eq2.2}).  
Thus, with the present quality of 
SNIa data, the statefinder parameters are, not surprisingly, 
no better at distinguishing between the models than a direct comparison 
with the SNIa data.  
We next look at simulated data to get an idea of how the situation will 
improve with future data sets.

\section{Future data sets}

We will now make an investigation of what an idealized SNIa survey 
can teach us about statefinder parameters and dark energy, 
following the procedure in Saini, Weller \& Bridle (2004). 

A SNAP-like satellite is expected to observe $\sim$2000 SN up to $z
\sim 1.7$. Dividing the interval $0 < x \leq 1.7$ into 50 bins, we
therefore expect $\sim$40 observations of SN in each bin. Empirically, 
SNIa are very good standard candles with a small dispersion in apparent 
magnitude $\sigma_{\rm mag}=0.15$, and there is no indication of 
redshift evolution.  The apparent magnitude is related to the luminosity 
distance through 
\begin{equation}
m(z) = {{\cal M}} + 5 \log D_{\rm L}(z),
\label{eq:magnitude}
\end{equation}
where ${{\cal M}} = M_0 + 5\log[H_0^{-1}{\rm Mpc}^{-1}]+25$.  
The quantity $M_0$ is the absolute magnitude of type Ia supernovae, and 
$D_{\rm L}(z) = H_0d_{\rm L}(z)$ is the Hubble constant free luminosity 
distance. The combination of absolute magnitude and the Hubble constant, 
${{\cal M}}$, can be calibrated by low-redshift supernovae (Hamuy et al. 1993; 
Perlmutter et al. 1999).  The dispersion in the magnitude, $\sigma_{\rm mag}$, 
is related to the uncertainty in the distance, $\sigma$, by 
\begin{equation}
\frac{\sigma}{d_{\rm L}(z)} = \frac{\ln 10}{5}\sigma_{\rm mag}, 
\label{eq:dlerr}
\end{equation}
and for $\sigma_{\rm mag}=0.15$, the relative error in the luminosity 
distance is $\sim 7$\%.  If we assume
that the $d_L$ we calculate for each $z$ value is the mean of
all $d_L$s in that particular bin, the errors reduce from 7\% to 
$0.07/\sqrt{40} \approx 0.01 = 1\%$.  We do not add noise to the 
simulated $d_{\rm L}$, and hence our results give the ensemble average 
of the parameters we fit to the simulated data sets. 

\subsection{A $\Lambda{\rm CDM}$ universe}
\begin{figure}[ht]
\begin{center}
\includegraphics[width=80mm,height=80mm]{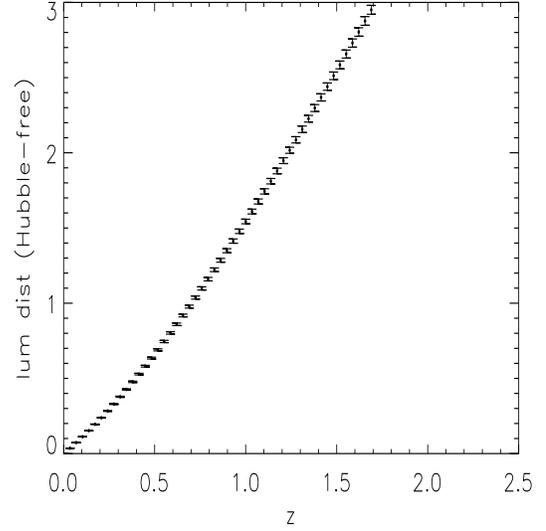}
\caption{Binned, simulated data set for a flat $\Lambda{\rm CDM}$ 
universe with $\Omega_{\rm m 0}=0.3$.    
The $1\sigma$ error bars are also shown.}
\label{fig:dldata1}
\end{center}
\end{figure}

\begin{figure}[ht]
\begin{center}
\includegraphics[width=80mm,height=80mm]{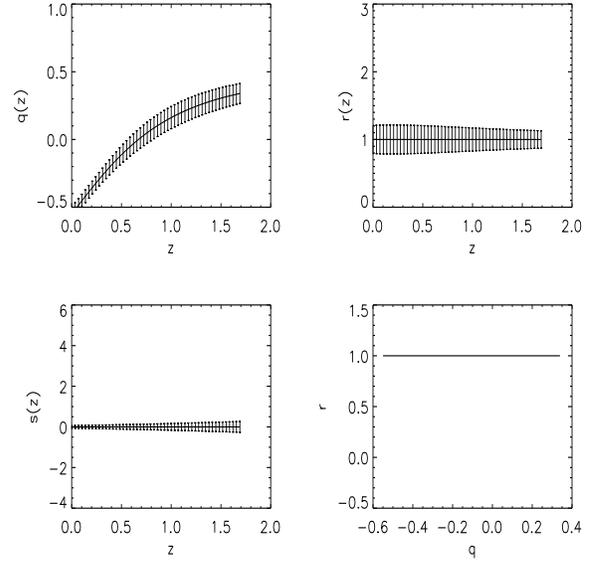}
\caption{The statefinder parameters and the deceleration parameter for the 
best-fitting reconstruction of the simulated data based on $\Lambda{\rm CDM}$, using the parametrization of Alam et al.  
The $1\sigma$ error bars are also shown.}
\label{fig:qrsplot1}
\end{center}
\end{figure}

We first simulate data based on a flat $\Lambda{\rm CDM}$ model with 
$\Omega_{\rm m0} = 0.3$, $h = 0.7$, giving the data points shown in 
figure \ref{fig:dldata1}. 
To this data set we first fit 
the quiessence model, the MPC, the GCG, and the parametrization of 
$H$ from equation (\ref{eq:eq2.2}).  Since all models reduce to $\Lambda{\rm CDM}$ 
for an appropriate choice of parameters, distinguishing between them 
based on the $\chi^2$ per degree of freedom alone is hard. 
Based on the best-fitting values and error bars on the parameters 
$A_0$, $A_1$, and $A_2$ in equation (\ref{eq:eq2.2}) we can  reconstruct 
the statefinder parameters from eqs. (\ref{eq:eq2.4}) -- (\ref{eq:eq2.6}).  
In figure \ref{fig:qrsplot1} we show the deceleration parameter and 
statefinder parameters reconstructed from the simulated data.  
The statefinders can be reconstructed quite well in this case, e.g. we 
see clearly that $r$ is equal to one, as it should for 
flat $\Lambda{\rm CDM}$.  
\begin{figure}[ht]
\begin{center}
\includegraphics[width=80mm,height=80mm]{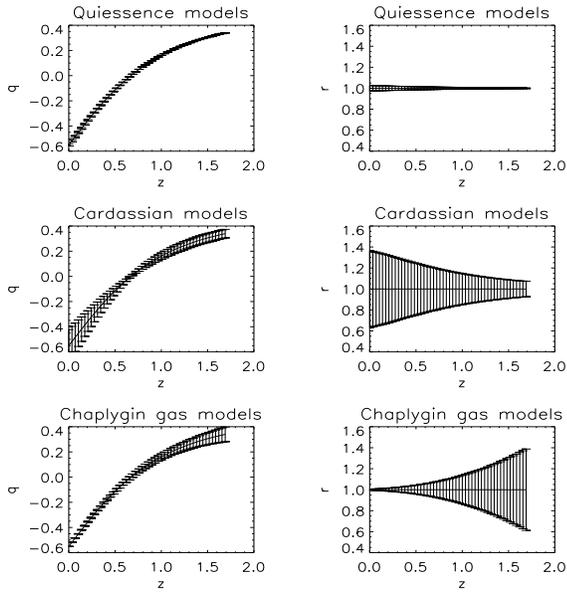}
\caption{The statefinder parameters for a selection of models, evaluated 
at the best-fitting values 
of their respective parameters to the simulated $\Lambda{\rm CDM}$ dataset,  
with errors 
included.}
\label{fig:qrsplot2}
\end{center}
\end{figure}    
In figure \ref{fig:qrsplot2} we show the statefinders for  
a selection of models, obtained by fitting their respective parameters 
to the data, and using the expressions for $q$ and $r$ for the respective 
models derived in earlier sections, e.g. equation (\ref{eq:eq1.46}) and 
(\ref{eq:eq1.47}) for the Chaplygin gas.   Since all models reduce  
to $\Lambda{\rm CDM}$ for the best-fitting parameters, their $q$ and $r$ 
values are also consistent with $\Lambda{\rm CDM}$.  Thus, if the 
dark energy really is LIVE, a SNAP-type experiment should be able to 
demonstrate this.

\subsection{A Chaplygin gas universe}

\begin{figure}[ht]
\begin{center}
\includegraphics[width=80mm,height=80mm]{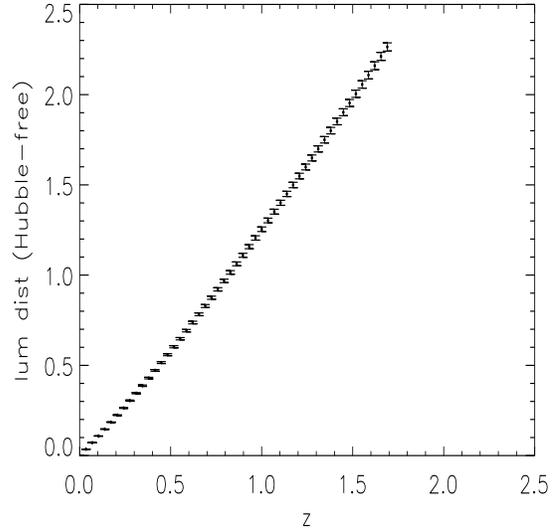}
\caption{Simulated luminosity distance-redshift relationship for the 
Generalized Chaplygin Gas with $A_{\rm s}=0.4$, $\alpha=0.7$.}
\label{fig:chapdlsim}
\end{center}
\end{figure}

We have also carried out the same reconstruction exercise using simulated 
data based on the GCG with $A_{\rm s} = 0.4$, $\alpha=0.7$, see 
figure \ref{fig:chapdlsim}.  
\begin{figure}[ht]
\begin{center}
\includegraphics[width=80mm,height=80mm]{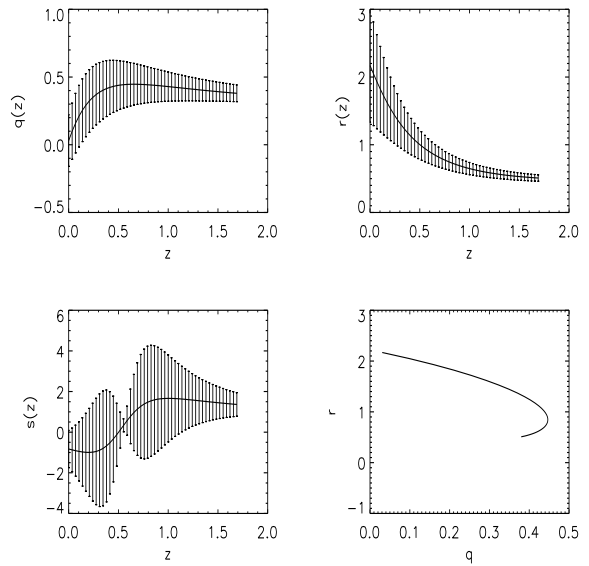}
\caption{The statefinder parameters and the deceleration parameter for the 
best-fitting reconstruction of the simulated data based on the GCG, 
using the parametrization of Alam et al.  
The $1\sigma$ error bars are also shown.}
\label{fig:qrsplotchap}
\end{center}
\end{figure}    
Figure \ref{fig:qrsplotchap} shows $q$ and $r$ reconstructed using the 
parametrization of $H$.  The same quantities for the models considered, 
based on their best-fitting parameters to the simulated data, are  
shown in figure \ref{fig:chapbasemodels}. 
\begin{figure}[ht]
\begin{center}
\includegraphics[width=80mm,height=80mm]{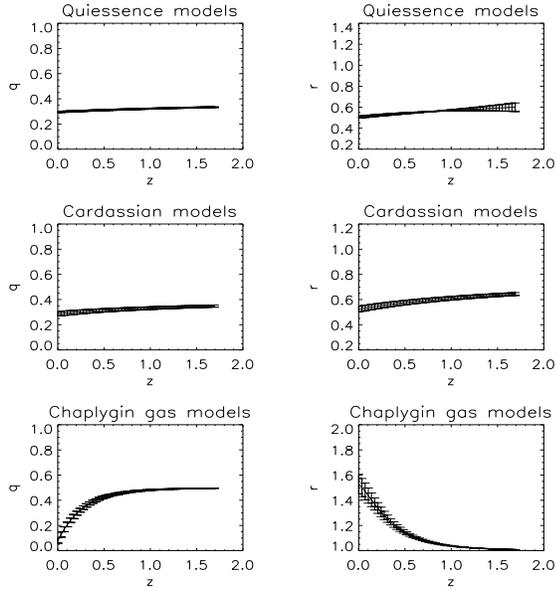}
\caption{The statefinder parameters for a selection of models, evaluated 
at the best-fitting values of their respective parameters to the simulated 
Chaplygin gas data set, with errors 
included.}
\label{fig:chapbasemodels}
\end{center}
\end{figure}  
For the Cardassian model, the best-fitting value for the parameter $n$, 
$n_{\rm bf}$, depends on the extent of the region over which we allow 
$n$ to vary.  
Extending this region to larger negative values for $n$ moves 
$n_{\rm bf}$ in the same direction.  However, the minimum $\chi^2$ 
value does not change significantly.  This is understandable, since 
$H(x)$ for the MPC model is insensitive to $n$ for large, negative 
values of $n$.  The quantities $r(x)$ and $q(x)$ also depend only 
weakly on the allowed range for $n$, whereas their error bars 
are sensitive to this parameter.  We chose to impose a prior $n > -1$, 
producing the results shown in figure \ref{fig:chapbasemodels}.  
The best-fitting values for $\psi$ and $n$ were, respectively, 
$0.06$ and $-0.94$.  
 
\begin{figure}[ht]
\begin{center}
\includegraphics[width=80mm,height=80mm]{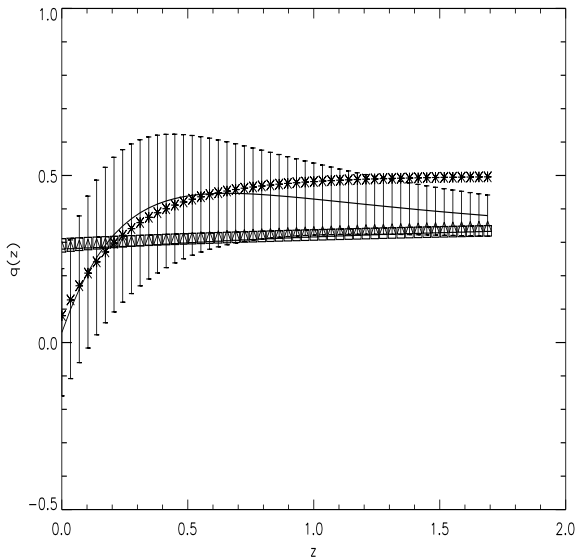}
\caption{Comparison of $q(z)$ extracted using the parametrized $H(z)$ with 
$q(z)$ for the various best-fitting models.  Error bars are only 
shown on the values extracted using the Alam et al. parametrization, 
but in the other cases they are roughly of the same size as the 
symbols.}
\label{fig:qfig}
\end{center}
\end{figure}

Figure \ref{fig:qfig} shows the deceleration parameter extracted from the Alam
et al. parametrization (full line), with $1\sigma$ error bars. 
Also plotted is the best fit $q(z)$ from the quiessence (squares), 
Cardassian (triangles) and
Chaplygin (asterisk) models. We note that the $q(z)$ from the Alam et al. 
parametrization has a somewhat deviating behaviour from the input model, 
especially at
larger $z$. Also, no model can be excluded on the basis of their predictions 
for $q(z)$

\begin{figure}[ht]
\begin{center}
\includegraphics[width=80mm,height=80mm]{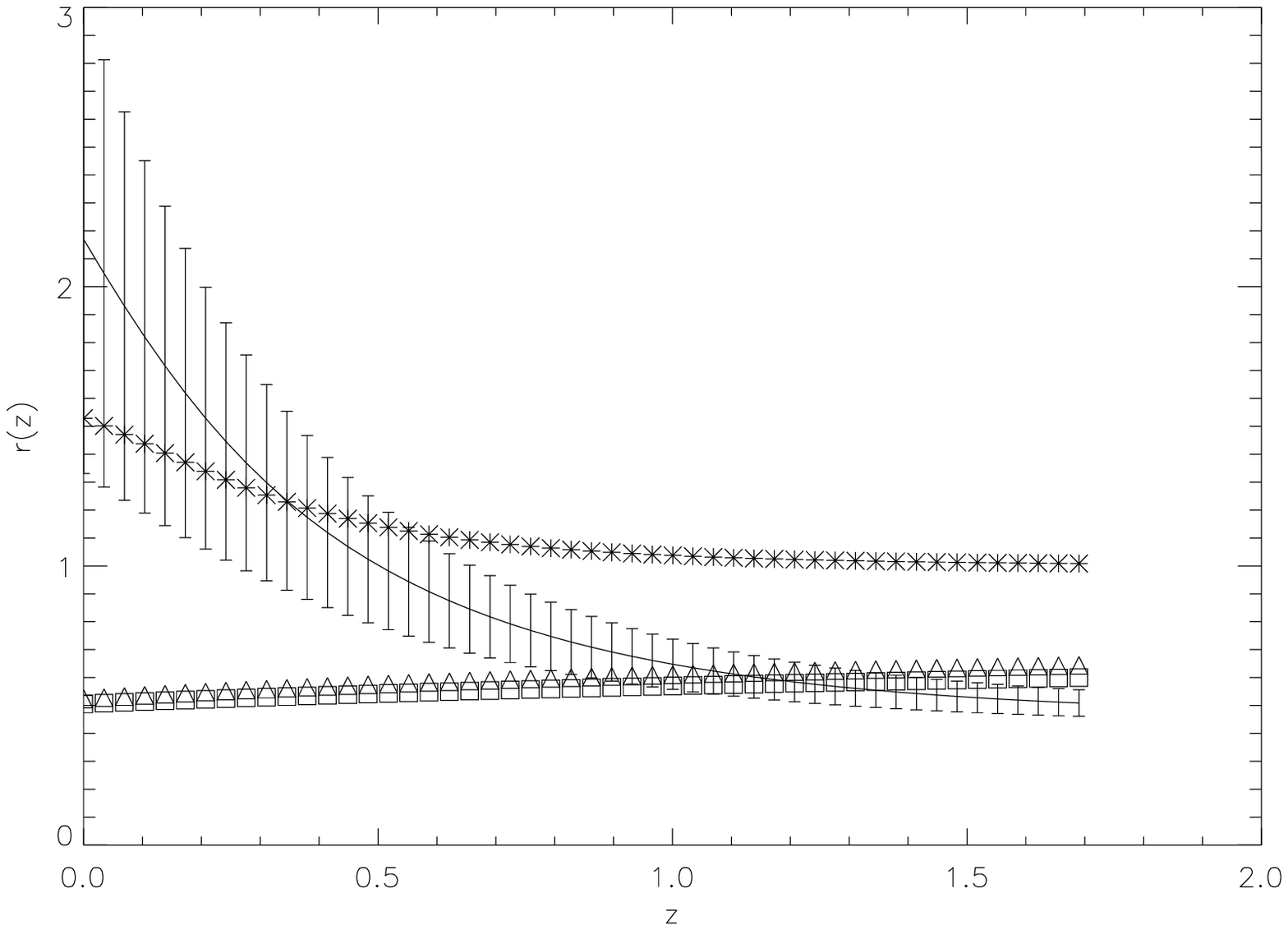}
\caption{Comparison of $r(z)$ extracted using the parametrized $H(z)$ 
with $r(z)$ for the various best-fitting models.  Error bars are only 
shown on the values extracted using the Alam et al. parametrization, 
but in the other cases they are roughly of the same size as the symbolsS.}
\label{fig:rfig}
\end{center}
\end{figure}

Figure \ref{fig:rfig} shows the same situation for the statefinder parameter
$r(z)$. Note again that for large $z$, the recovered statefinder from the
Alam et al. parametrization does not correspond well with the input model. As
with the case for $q(z)$, the quiessence and Cardassian models follow each other
closely. These, however, do not agree with the input model for
low values of $z$ (similar to the case for $q(z)$ they diverge for low $z$). 
Comparing the statefinder $r$ for the quiessence and Cardassian models 
with that of the input GCG model, indicates that, not surprisingly, 
neither of them is a good fit to the data.

This exercise indicates that the statefinders can potentially 
distinguish between dark energy models, if $q$, $r$ and $s$ can 
be extracted from the data in a reliable, model-independent way.  
However, the fact that $r$ extracted from the simulated 
data using the Alam et al. parametrization does not agree well with 
the true $r$ of the underlying model for $z>0.5$, indicates that one 
needs a better parametrization in order to use statefinder parameters 
as empirical discriminators between dark energy models.

\section{Conclusions}

We have investigated the statefinder parameters as a means of 
comparing dark energy models.  As a theoretical tool, they are 
very useful for visualizing the behaviour of different 
dark energy models.  Provided they can be extracted from 
the data in a reliable, model-independent way, they can give 
a first insight into the type of model which is likely to describe 
the data.  However, SNIa data of quality far superior to those 
presently available are needed in order to distinguish between 
the different models.  And even with SNAP-quality data, there 
may be difficulties in distinguishing between models based on 
the statefinder parameters alone.  Furthermore, the parametrization of 
$H(z)$ used here and in Alam et al. (2003) is probably not 
optimal, as shown in section 4.2.   The same conclusion was 
reached in a recent investigation by J\"{o}nsson et al. (2004), 
where they considered reconstruction of the equation of state 
$w(x)$ from SNIa data using equation (\ref{eq:eq2.65}).   
They found that this parametrization forces the behaviour of  
$w(x)$ onto a specific set of tracks, and may thus give spurious 
evidence for redshift evolution of the equation of state.  Although  
this conclusion has been contested by Alam et al. (2004), it is clear 
that finding a parametrization which is sufficiently general, and at 
the same time with reasonably few parameters is an important task 
for future work.

\begin{acknowledgements}
We acknowledge support from the Research Council of Norway (NFR) through 
funding of the project `Shedding Light on Dark Energy'.    
The authors wish to thank H\aa vard Alnes for interesting discussions.

\end{acknowledgements}


\begin{thebibliography}{}

\bibitem[Alam \& Sahni (2002)]{alam0} Alam, U., Sahni, V. 2002, astro-ph/0209443
\bibitem[Alam et al. (2003)]{alam} Alam, U., Sahni, V., Saini, T. D., Starobinsky, A. A. 2003, MNRAS, 344, 1057
\bibitem[Alam et al. (2004)]{alam2} Alam, U., Sahni, V., Saini, T. D., Starobinsky A. A. 2004, astro-ph/0406672
\bibitem[Bento, Bertolami \& Sen (2002)] Bento, M. C., Bertolami, A. A., 
Sen, A. A., 2002, Phys. Rev. D, 66, 043507
\bibitem[Bertolami et al. (2004)]{bertolami} Bertolami, O., Sen, A. A., Sen, S., Silva, P. T., 2004, MNRAS (in press), astro-ph/0402387
\bibitem[Bilic, Tupper \& Viollier (2002)]{chaplygin2} Bilic, N., Tupper, G. G., Viollier, R. 2002, Phys. Lett. B, 535, 17
\bibitem[Bucher \& Spergel (1999)]{bucher} Bucher, M., Spergel, D. 1999, Phys. Rev. D, 60, 043505
\bibitem[Caldwell (2002)]{phantom} Caldwell, R. R. 2002, Phys. Lett. B, 545, 23
\bibitem[Caldwell \& Kamionkowski (2004)]{caldwell} Caldwell, R. R., Kamionkowski M. 2004, preprint astro-ph/0403003
\bibitem[Chiba \& Nakamura (1998)]{chiba} Chiba ,T., Nakamura T. 1998, Prog. Theor. Phys., 100, 1077 
\bibitem[Deffayet (2001)]{deffayet1} Deffayet, C. 2001, Phys. Lett. B, 502, 199
\bibitem[Deffayet, Dvali \& Gabadadze 2002]{deffayet2} Deffayet, C, Dvali, G., Gabadadze, G. 2002, Phys. Rev. D, 65, 044023
\bibitem[Dvalie, Gabadadze \& Porrati (2000)]{dvali1} Dvali, G., Gabadadze, G., Porrati, M. 2000, Phys. Lett. B, 485, 208
\bibitem[Efstathiou et al. (2002)]{efstathiou} Efstathiou, G., et al. 2002, MNRAS, 330, L29
\bibitem[Eichler (1996)]{eichler} Eichler, D. 1996, ApJ, 468, 75
\bibitem[Freese \& Lewis (2002)]{freese1} Freese, K., Lewis, M. 2002, Phys. Lett. B, 540, 1 
\bibitem[Gondolo \& Freese (2003)]{gondolo} Gondolo, P., Freese, K. 2003, 
Phys. Rev. D, 68, 063509
\bibitem[Gong (2004)]{gong} Gong, Y. 2004, preprint astro-ph/0405446
\bibitem[Gorini, Kamenshchik \& Moschella (2003)]{gorini} Gorini, V., 
Kamenshchik, A., Moschella U. 2003, Phys. Rev. D, 67, 063509
\bibitem[Hamuy et al. (1993)]{hamuy} Hamuy, M., et al. 1993, ApJ, 106, 2392
\bibitem[Jeffreys (1961)]{jeffreys} Jeffreys, H. 1961, {\it Theory of probability}, 3rd ed., Oxford University Press
\bibitem[J\"{o}nsson et al. (2004)]{jonsson} J\"{o}nsson, J., Goobar, A., 
Amanullah, R., Bergstr\"{o}m, L. 2004, preprint astro-ph/0404468
\bibitem[Kamenshchik, Moschella \& Pasquier (2001)]{chaplygin1} Kamenshchick, A., Moschella, U., Pasquier, V. 2001, Phys. Lett. B, 511, 265 
\bibitem[Liddle (2004)]{liddle} Liddle, A. 2004, preprint astro-ph/0401198
\bibitem[Padmanabhan (2002)]{paddy1} Padmanabhan, T. 2002, Phys. Rev. D, 66, 021301
\bibitem[Padmanabhan \& Choudhury (2003)]{paddy2} Padmanabhan, T., Choudhury, R. 2003, MNRAS, 344, 823
\bibitem[Peebles \& Ratra (1988)]{peebles} Peebles, P. J. E., Ratra, B. 1988, ApJ, 325, L17
\bibitem[Perlmutter et al. (1999)]{perlmutter} Perlmutter, S., et al. 1999, Ap. J., 517, 565
\bibitem[Riess et al. (1998)]{riess} Riess, A. G., et al. 1998, AJ, 116, 1009 
\bibitem[Riess et al. (2004)]{riess2} Riess, A. G., et al. 2004, ApJ, 607, 655
\bibitem[Sahni \& Shtanov (2003)]{shtanov} Sahni, V., Shtanov, Y. 2003, JCAP, 0311, 014 
\bibitem[Sahni et al. (2003)]{sahni} Sahni, V., Saini, T. D., Starobinsky, A. A., Alam, U. 2003, JETP Lett., 77, 201
\bibitem[Saini, Weller \& Bridle (2004)]{saini} Saini, T. D., Weller, J., Bridle S. L. 2004, MNRAS, 348, 603
\bibitem[Schwarz (1978)]{schwarz} Schwarz, G. 1978, Annals of Statistics, 5, 461
\bibitem[Solheim (1966)]{solheim} Solheim, J.-E. 1966, MNRAS, 133, 32
\bibitem[Tegmark et al. (2003)]{tegmark1} Tegmark, M., et al. 2004, Phys. Rev. D, 69, 103501
\bibitem[Visser (2003)]{visser} Visser, M. 2004, Class. Quant. Grav., 21, 2603
\bibitem[Wetterich (1988)]{wetterich} Wetterich, C. 1988, Nucl. Phys. B, 302, 668 



\end{thebibliography}
\end{document}